\documentclass[prd,twocolumn,
showpacs,preprintnumbers,nofootinbib]{revtex4}
\usepackage[dvips]{graphicx}
\usepackage{enumerate}
\usepackage{amsmath,amssymb}
\usepackage{mathrsfs}
\usepackage[dvips]{graphicx}
\usepackage{bm}

\begin{document}
\preprint{CHIBA-EP-196, 2012}

\title{Physical unitarity  for a massive Yang-Mills theory without the Higgs field:
\\  A perturbative treatment
}

\author{Kei-Ichi Kondo}
\email[]{kondok@faculty.chiba-u.jp}

\author{Kenta Suzuki}

\author{Hitoshi Fukamachi}

\author{Shogo Nishino}

\author{Toru Shinohara}
%\email[]{sinohara@graduate.chiba-u.jp}

\affiliation{
$^1$Department of Physics, Graduate School of Science, Chiba University,
Chiba 263-8522, Japan
}

\begin{abstract}
In a series of papers, we examine the physical unitarity in a massive Yang-Mills theory without the Higgs field in which the color gauge symmetry is not spontaneously broken and kept intact. 
For this purpose, we use a new framework proposed in the previous paper Kondo [arXiv:1208.3521] based on a nonperturbative construction of a non-Abelian field describing a massive spin-one vector boson field, which enables us to perform the perturbative and nonperturbative studies on the physical unitarity. 
In this paper, we present a new perturbative treatment for the physical unitarity after giving the general properties of the massive Yang-Mills theory. Then  we reproduce  the violation of physical unitarity in a transparent way.
This paper is a preliminary work to the subsequent papers in which we present a nonperturbative framework to propose a possible scenario of restoring the physical unitarity in the Curci-Ferrari model. 

\end{abstract}


\maketitle

%%%%%%%%%%%%%%%%%%%%%%%%%%%%%%%%%%%%%%%%%%%%%%%%%%%%%%%%%%%%%%%%%%%%%%%%
%%%%%%%%%%%%%%%%%%%%%%%%%%%%%%%%%%%%%%%%%%%%%%%%%%%%%%%%%%%%%%%%%%%%%%%%
\section{Introduction}
%%%%%%%%%%%%%%%%%%%%%%%%%%%%%%%%%%%%%%%%%%%%%%%%%%%%%%%%%%%%%%%%%%%%%%%%
%%%%%%%%%%%%%%%%%%%%%%%%%%%%%%%%%%%%%%%%%%%%%%%%%%%%%%%%%%%%%%%%%%%%%%%%

In the Yang-Mills theory \cite{YM54} and quantum chromodynamics (QCD) for strong interactions,  both the renormalizability and the physical unitarity are satisfied, as demonstrated first in \cite{tHooft71}.
%In these theories, the Yang-Mills field is assumed to be massless at least in the original Lagrangian.  
Moreover, it is also known that the massive Yang-Mills theory satisfies both the renormalizability and the physical unitarity \cite{tHooft71b}, if the local gauge invariance is spontaneously broken by the Higgs field \cite{Higgs66} and the gauge field acquires the mass through the Higgs mechanism by absorbing the Nambu-Goldstone particle associated with the spontaneous symmetry breakdown.
%['t Hooft, 1971]
%Nucl. Phys. B35, 167 (1971).
%and 
%[Lee and Zinn-Justin, 1972].
In other words, both the renormalizability and the physical unitarity survive the spontaneous breaking of the gauge symmetry.

%In the Glashow-Weinberg-Salam (GWS) model of the electro-weak interactions, on the other hand, the gauge bosons mediating the weak interactions must be massive, since the weak interactions are of the short-range.  Therefore, how to describe the massive gauge boson is an important issue to be clarified in quantum gauge field theories.  
%In the GWS model, the gauge bosons acquire  their masses through the Higgs mechanics, by absorbing the would-be Nambu-Goldstone particles associated with the spontaneous breaking of the original gauge symmetry $SU(2)_L \times U(1)_Y$ to the subgroup $U(1)_{EM}$. In fact, such  massive gauge bosons $W^+, W^-, Z^0$ have been discovered in the mid-1980 by high-energy experiments. 
%For  the gauge bosons to be massive through the Higgs mechanism, the existence of the Higgs scalar field is indispensable, although the Higgs particle has not yet been detected.  
%People expect that the Higgs particle   will be sooner or later discovered in experiments, inspired by the success of the standard model for elementary particles.   

It is a long-standing problem \cite{DV70,SF70,Boulware70,CF76,CF76b,FMTY81,Ojima82,BSNW96,DTT88,RRA04,BFQ} to clarify whether it is possible or not to construct a massive Yang-Mills model blessed with both the physical unitarity and the renormalizability without the Higgs fields, in which the local gauge symmetry is not spontaneously broken.
Here the Lagrangian is assumed to be written in polynomials of the fields (we exclude the nonpolynomial type \cite{FMTY81} from our discussions).
We are anxious to find such a model for understanding the mass gap and confinement caused by the strong interactions  \cite{Cornwall82,decoupling,decoupling-lattice,TW11,scaling}, since the Higgs field does not exist and the color gauge symmetry is kept intact in QCD. 
%However, a fact that the strong interactions are of the short-range has not yet been understood, 
Indeed, there are continued attempts to look for an alternative way to describe massive non-Abelian gauge fields without the Higgs field \cite{DV70,SF70,Boulware70,CF76,CF76b,FMTY81,Ojima82,BSNW96,DTT88,RRA04}. 
However, all these efforts were unsuccessful in coping with both renormalizability and unitarity very well:
In all the models proposed so far for the massive Yang-Mills theory without the Higgs fields, it seems that  the renormalizability and the physical unitarity are not compatible with each other,
 although there are some models which satisfy either the renormalizability or the physical unitarity.
See \cite{DTT88,RRA04} for reviews and \cite{BFQ} for later developments.

%apart from  the shortcomings in the Higgs approach, e.g., the indeterminate masses and couplings of the Higgs particles themselves.

For this purpose, we start once again from the Curci-Ferrari (CF) model \cite{CF76}, which is a massive extension of the massless Yang-Mills theory in the most general renormalizable gauge having both the Becchi-Rouet-Stora-Tyutin (BRST) and anti-BRST symmetries \cite{Baulieu85}.
In preceding studies for the CF model \cite{CF76,CF76b,Ojima82,BSNW96}, the CF model is proved to be renormalizable, whereas the CF model has been concluded to violate physical unitary \cite{CF76b,Ojima82,BSNW96}.
% when restricted to the physical subspace specified by the Kugo-Ojima (KO) subsidiary condition \cite{KO78}. 
However, the preceding studies are restricted to considerations in the perturbation theory. 
We need a nonperturbative framework to draw a definite conclusion to this issue.

In a previous paper \cite{Kondo12}, therefore, we have presented a nonperturbative construction of a massive Yang-Mills field  $\mathscr{K}_\mu$ which  describes a non-Abelian massive spin-one vector boson with the correct physical degrees of freedom without the Higgs field \cite{Higgs66}. 
This is achieved by finding a nonlinear but local transformation from the original fields in the CF model to the physical massive vector field $\mathscr{K}_\mu$ which is invariant under the modified BRST and anti-BRST transformation. 
As an application, we have written down a local mass term for the Yang-Mills field and a dimension-two condensate, which are exactly invariant under the modified BRST transformation,  Lorentz transformation and  color rotation. 
The resulting massive Yang-Mills model is regarded as a low-energy effective theory of QCD, which enables us to understand the decoupling solution \cite{decoupling}  characterizing   the deep infrared regime responsible  for color confinement \cite{KO79,Gribov78}. 

In a series of papers, we give the perturbative and nonperturbative studies on the physical unitarity \cite{Kondo12a} in the massive Yang-Mills theory constructed in the previous paper \cite{Kondo12}.
In the ordinary massless Yang-Mills theory, the physical unitarity is a first step of understanding color confinement \cite{KO79}: In the intermediate state, the contributions from the unphysical gauge modes, i.e., the longitudinal and scalar modes are exactly canceled by those of the ghost and antighost, which is a special case of  the quartet mechanism \cite{KO78}.
We clarify how the situation changes in the massive case. 
Moreover, we clarify the reason for failures of the preceding attempts from our viewpoint.  

This paper is the first  of the planned papers for discussing the perturbative and nonperturbative physical unitarity in the massive Yang-Mills theory without the Higgs field.
In this paper we present a new perturbative treatment for the physical unitarity after reviewing the general properties of the massive Yang-Mills theory. Then  we reproduce  the violation of physical unitarity in a transparent way.
In subsequent papers, we present a nonperturbative framework to discuss a possible scenario of restoring the physical unitarity in the massive Yang-Mills theory.

Finally, we mention the difference between the unitarity and physical unitarity of the scattering matrix from our point of view. 
For the tree-level scattering amplitude between two longitudinally polarized vector bosons, it is known \cite{PS95,CLT73,LQT77} that the scattering probability as a function of the energy $E$ becomes greater than 1 above a critical value $E_c$, since the amplitude grows with the energy $E$ like $g^2E^2/M^2$ where $M$ is the mass of the vector boson and $g$ is the coupling constant for the self-interactions among vector bosons.
This implies that the perturbative unitarity breaks down in   high-energy  $E \ge E_c$.
Therefore, for the perturbative  unitarity to be satisfied, the energy must be restricted to low-energy   $E < E_c$, which is called the {unitarity bound}. 
(The violation of the unitarity condition for the scattering amplitude in high energy is understood  from the Nambu-Goldstone equivalence theorem \cite{CLT73} and the low-energy theorem.)
In the Higgs sector of weak interactions in the standard model, the Higgs particle exists and the exchange of the Higgs particle affects the amplitude so that the amplitude approaches a constant at energies far above the Higgs pole. 
Consequently, the Higgs mass must be less than an upper bound.
If such new physical degrees of freedom do not exist, this behavior is not modified and the unitarity violation in high energy cannot be avoided in the massive Yang-Mills theory, since   the Nakanishi-Lautrup (NL) auxiliary field can be integrated out and the ghosts can play no role in the tree-level amplitude. 
In our works, we regard the CF model as a low-energy effective theory of the Yang-Mills theory to be valid in the region $E < E_c$  for discussing  color confinement.
We restrict our examination on the physical unitarity  to a sufficiently  low-energy region below a few GeV  to evade the unitarity violation and consider only the physical unitarity, i.e., unphysical mode cancellation in our papers. 
Therefore, the well-known fact about the unitarity violation in the above does not contradict  our research on the physical unitarity.

%If we succeed to describe the massive gauge bosons in coping with both renormalizability and unitarity, we have the following advantages. 

%Currently, the Higgs mechanism \cite{Higgs66} is widely accepted as a unique way for providing the masses for gauge bosons in quantum field theories, because it is the only one established method  which enables one to  maintain both renormalizability and unitarity \cite{tHooft71} in massive non-Abelian gauge field theories \cite{YM54}.   
%Indeed, the Higgs particle becomes an indispensable ingredient in the unified model of electro-weak interactions, although it has not yet been  detected, but it is to be discovered in the near future.

%In contrast to these results, we show that the CF model fulfills the physical unitarity, if the massive vector field is correctly identified. 

This paper is organized as follows.

In section II, we introduce a massive Yang-Mills theory without the Higgs field and define the CF model as a special case. 
The CF model is not invariant under the usual BRST and anti-BRST transformations.  
However, the CF model can be made invariant by modifying the BRST and anti-BRST transformations.  
The cost of introducing the mass term is the violation of nilpotency of the modified BRST and anti-BRST transformations. 
We point out an important fact that even the modified BRST (anti-BRST)-invariant quantity depends on a parameter $\beta$ in the $M \not= 0$ case.  
This should be compared with the  $M=0$  case, in which $\beta$ is a gauge-fixing parameter and the BRST-invariant quantity does not depend on $\beta$, which means that the physics does not depend on $\beta$ in the $M=0$ case. 
This is not the case for $M \not= 0$.

In section III, we summarize the result obtained in the previous paper \cite{Kondo12} on a nonperturbative construction of a non-Abelian massive Yang-Mills field $\mathscr{K}_\mu$ 
under the requirements which guarantee (i) the modified BRST (and anti-BRST) invariance, (ii) correct degrees of freedom for describing a massive spin-one particle, and (iii) the expected transformation rule under  color rotation. 
We write down the massive vector field explicitly in terms of the original Yang-Mills field, the Faddeev-Popov  (FP) ghost field, antighost field and the  NL  field in the CF model.

In section IV, we give a perturbative framework of the CF model in terms of the new field variable $\mathscr{K}_\mu$.  We give the Feynman rules up to the order $g$. 

In section V, we check the physical unitarity in the massless Yang-Mills theory.
 Using a simple example, it is demonstrated in the lowest order of perturbation theory  that the physical unitarity follows from the cancellation among unphysical modes: the  longitudinal and scalar modes of the Yang-Mills field together with the FP ghost and antighost.    

In section VI, we review a conventional argument for the violation of physical unitarity in the massive Yang-Mills theory without the Higgs field.  
Using a simple example corresponding to the previous section, we show that the violation of physical unitarity follows from the incomplete cancellation among unphysical modes: the scalar mode with the FP ghost and antighost.

In section VII, we begin with a new analysis on the physical unitarity of the CF model based on a novel framework using the field $\mathscr{K}_\mu$ given in section III.
In this section, we give a new perturbative analysis using the result of section V. 
We confirm that the physical unitarity is indeed violated in the CF model in the framework of the perturbation theory in the coupling constant. 
It is easily seen that the violation of physical unitarity follows from the incomplete cancellation among unphysical modes: the NL field (corresponding to the scalar mode) with the FP ghost and antighost.    
We discuss how to avoid the violation of physical unitarity within the perturbative framework.

%In section IX and X, we give a nonperturbative analysis on physical unitarity.  
%First, in section IX, we clarify the relationship between the scalar mode of the Yang-Mills field and the NL field on mass shell. 
%In the perturbative framework, we see that they are equivalent  on mass shell, namely,   on mass shell the scalar mode of the Yang-Mills field is identified with the NL field. This is an origin of perturbative violation of physical unitarity in the massive case. 
%This is avoided in a nonperturbative way, which leads to a possibility to recover the physical unitarity in the massive case, as pointed out in section X. 

%In section XI, we give some remarks related to violation of nilpotency of the modified BRST transformation, physical degrees of freedom in the massless and massive cases, and the relationship between the renormalizability and physical unitarity. 

In the final section, we summarize the results and mention the perspective on the next work.

In Appendix A,  we calculate the Jacobian associated with the change of variables from the original CF model to the new theory written in terms of new variables. 
In Appendix B,  the Feynman rules are given up to the next order $g^2$, with which we supplement  the results of section V.

%\newpage
%%%%%%%%%%%%%%%%%%%%%%%%%%%%%%%%%%%%%%%%%%%%%%%%%%%%%%%%%%%%%%%%%%%%%%%%
%%%%%%%%%%%%%%%%%%%%%%%%%%%%%%%%%%%%%%%%%%%%%%%%%%%%%%%%%%%%%%%%%%%%%%%%
\section{The Curci-Ferrari model and the modified BRST transformation}
%%%%%%%%%%%%%%%%%%%%%%%%%%%%%%%%%%%%%%%%%%%%%%%%%%%%%%%%%%%%%%%%%%%%%%%%
%%%%%%%%%%%%%%%%%%%%%%%%%%%%%%%%%%%%%%%%%%%%%%%%%%%%%%%%%%%%%%%%%%%%%%%%

In order to look for a candidate of the massive Yang-Mills theory without the Higgs field, we start from the usual massless Yang-Mills theory in the most general Lorentz gauge formulated in a manifestly Lorentz covariant way.
The total Lagrangian density is written in terms of the Yang-Mills field $\mathscr{A}_\mu$, the  FP ghost field $\mathscr{C}$, the antighost field $\bar{\mathscr{C}}$ and the  NL field $\mathscr{N}$.
As a candidate of the massive Yang-Mills theory without the Higgs field,
we add the ``mass term'' $\mathscr{L}_m$:
\begin{subequations}
	\begin{align}
		\mathscr{L}^{\rm{tot}}_{m\rm{YM}} =& \mathscr{L}_{\rm{YM}} + \mathscr{L}_{\rm{GF+FP}} + \mathscr{L}_{m} , 
\\
		\mathscr{L}_{\rm{YM}}  =& - \frac{1}{4} \mathscr{F}_{\mu \nu} \cdot \mathscr{F}^{\mu \nu} , 
   \\
		\mathscr{L}_{\rm{GF+FP}}  =& \frac{\alpha}{2} \mathscr{N} \cdot \mathscr{N}  + \frac{\beta}{2} \mathscr{N} \cdot \mathscr{N} 
  \nonumber\\
        & + \mathscr{N} \cdot \partial^{\mu} \mathscr{A}_{\mu} 
		- \frac{\beta}{2} g \mathscr{N} \cdot (i \bar{\mathscr{C}} \times \mathscr{C}) 
  \nonumber\\
		& + i \bar{\mathscr{C}} \cdot \partial^{\mu} \mathscr{D}_{\mu}[\mathscr{A}] \mathscr{C}
  \nonumber\\
		&+ \frac{\beta}{4} g^2 (i \bar{\mathscr{C}} \times \mathscr{C}) \cdot (i \bar{\mathscr{C}} \times \mathscr{C}) 
\nonumber\\
		 =& \mathscr{N} \cdot \partial^{\mu} \mathscr{A}_{\mu} + i \bar{\mathscr{C}} \cdot \partial^{\mu} \mathscr{D}_{\mu}[\mathscr{A}] \mathscr{C}
  \nonumber\\
		& + \frac{\beta}{4} ( \bar{\mathscr{N}} \cdot \bar{\mathscr{N}} + \mathscr{N} \cdot \mathscr{N}) 
		+ \frac{\alpha}{2} \mathscr{N} \cdot \mathscr{N}
, 
   \\
		\mathscr{L}_{m}  =& \frac{1}{2} M^2 \mathscr{A}_{\mu} \cdot \mathscr{A}^{\mu} + \beta M^2 i \bar{\mathscr{C}} \cdot \mathscr{C} , 
	\end{align}
%		\label{mYM1}
\end{subequations}
where $\alpha$ and $\beta$ are parameters corresponding to  the gauge-fixing parameters in the $M \rightarrow 0$ limit, 
$
 \mathscr{D}_{\mu}[\mathscr{A}] \mathscr{C}(x) 
%:= \partial_{\mu} \mathscr{C}(x) - ig [\mathscr{A}(x), \mathscr{C}(x)] 
:=  \partial_{\mu}\mathscr{C}(x) + g \mathscr{A}(x) \times \mathscr{C}(x)  
$,
 and 
\begin{equation}
\bar{\mathscr{N}} :=-\mathscr{N}+gi\bar{\mathscr{C}} \times \mathscr{C} .
\end{equation}

The $\alpha=0$ case is the  CF  model with the coupling constant $g$, the mass parameter $M$ and the parameter $\beta$.
In the Abelian limit with  vanishing structure constants $f^{ABC}=0$, the FP ghosts decouple and the CF model reduces to the Nakanishi model \cite{Nakanishi72}.

In what follows, we restrict our considerations to the $\alpha=0$ case. 
In the $\alpha=0$ case,   $\mathscr{L}_{\rm YM} + \mathscr{L}_{\rm GF+FP}$ 
is constructed so as to be invariant 
under both the usual BRST transformation: 
	\begin{align}
		\begin{cases}
			{\boldsymbol \delta}  \mathscr{A}_{\mu}(x) = \mathscr{D}_{\mu}[\mathscr{A}] \mathscr{C}(x)  \\
			{\boldsymbol \delta}  \mathscr{C}(x) = -\frac{g}{2} \mathscr{C}(x) \times \mathscr{C}(x)   \\
			{\boldsymbol \delta}  \bar{\mathscr{C}}(x) = i \mathscr{N}(x)   \\
			{\boldsymbol \delta}  \mathscr{N}(x) = 0  \\
		\end{cases} ,
		\label{BRST}
	\end{align}
and anti-BRST transformation: 
\begin{align}
\begin{cases}
			\bar{\boldsymbol \delta}  \mathscr{A}_{\mu}(x) = \mathscr{D}_{\mu}[\mathscr{A}] \bar{\mathscr{C}}(x) \\
			\bar{\boldsymbol \delta}  \bar{\mathscr{C}}(x) %= ig \{C(x), C(x) \} (
= -\frac{g}{2} \bar{\mathscr{C}}(x) \times \bar{\mathscr{C}}(x)  \\
			\bar{\boldsymbol \delta}   \mathscr{C}(x) = i \bar{\mathscr{N}}(x) \\
			\bar{\boldsymbol \delta}  \bar{\mathscr{N}}(x) = 0
\end{cases} 
 .
\end{align}
Indeed, it is checked that 
\begin{align}
		{\boldsymbol \delta}  \mathscr{L}_{\rm{YM}} = 0 ,
\quad
{\boldsymbol \delta}  \mathscr{L}_{\rm{GF+FP}} = 0 ,
\\
		\bar{\boldsymbol \delta}  \mathscr{L}_{\rm{YM}} = 0 ,
\quad
\bar{\boldsymbol \delta}  \mathscr{L}_{\rm{GF+FP}} = 0 .
	\end{align}

This is not the case for the mass term $\mathscr{L}_{m}$, i.e., 
	\begin{align}
		{\boldsymbol \delta} \mathscr{L}_{m}  \ne 0 .
	\end{align}
Even in the presence of the mass term $\mathscr{L}_{m}$, however, the  total Lagrangian $\mathscr{L}_{\rm mYM}^{\rm tot}$ can be made invariant by modifying the BRST transformation \cite{CF76}:
$\delta_{\rm BRST}'=\lambda {\boldsymbol \delta}'$ with a Grassmannian number $\lambda$ and
\begin{align}
\begin{cases}
			{\boldsymbol \delta}' \mathscr{A}_{\mu}(x) =  \mathscr{D}_{\mu}[\mathscr{A}] \mathscr{C}(x) \\
			{\boldsymbol \delta}' \mathscr{C}(x) %= ig \{C(x), C(x) \} (
=  -\frac{g}{2} \mathscr{C}(x) \times \mathscr{C}(x)  \\
			{\boldsymbol \delta}' \bar{\mathscr{C}}(x) =  i \mathscr{N}(x) \\
			{\boldsymbol \delta}' \mathscr{N}(x) =  M^2 \mathscr{C}(x)  
\end{cases} 
 .
\end{align}
The modified BRST transformation deforms the BRST transformation of the NL field and reduces to the usual BRST transformation in the limit $M \rightarrow 0$.
It should be remarked that 
  ${\boldsymbol \delta}' \mathscr{L}^{\rm tot}_{\rm mYM}=0$ follows from
	\begin{align}
		0 = {\boldsymbol \delta}' (\mathscr{L}_{\rm GF+FP} + \mathscr{L}_{m}) ,
		\label{req}
	\end{align}	
while
 	\begin{equation}
{\boldsymbol \delta}' \mathscr{L}_{m} \ne 0, 
\quad
{\boldsymbol \delta}' \mathscr{L}_{\rm GF+FP} \ne 0.
	\end{equation}

Similarly, the total action is  invariant under a modified anti-BRST transformation $\bar{{\boldsymbol \delta}}'$
defined by 
\begin{align}
\begin{cases}
			\bar{\boldsymbol \delta}' \mathscr{A}_{\mu}(x) = \mathscr{D}_{\mu}[\mathscr{A}] \bar{\mathscr{C}}(x) \\
			\bar{\boldsymbol \delta}' \bar{\mathscr{C}}(x) %= ig \{C(x), C(x) \} (
= -\frac{g}{2} \bar{\mathscr{C}}(x) \times \bar{\mathscr{C}}(x)  \\
			\bar{\boldsymbol \delta}'  \mathscr{C}(x) = i \bar{\mathscr{N}}(x) \\
			\bar{\boldsymbol \delta}' \bar{\mathscr{N}}(x) = - M^2 \bar{\mathscr{C}}(x)  
\end{cases} 
 ,
\end{align}
which reduces to the usual anti-BRST transformation in the limit $M \to 0$.
It is sometimes useful to give another form:
\begin{align}
 {\boldsymbol \delta}' \mathscr{\bar N}(x) =& g \mathscr{\bar N}(x) \times \mathscr{C}(x) - M^2 \mathscr{C}(x) ,
\nonumber\\
\bar{{\boldsymbol \delta}}' \mathscr{N}(x) =& g \mathscr{N}(x) \times \bar{\mathscr{C}}(x) + M^2 \bar{\mathscr{C}}(x) .
\end{align}

	Moreover, the path-integral integration measure  
$\mathcal{D} \mathscr{A} \mathcal{D} \mathscr{C} \mathcal{D} \bar{\mathscr{C}} \mathcal{D} \mathscr{N}$
is invariant under the modified BRST transformation. 
Indeed, it has been shown in \cite{Kondo12} that the Jacobian associated to the change of integration variables $\Phi(x) \to \Phi'(x) =\Phi(x)+ \lambda {\boldsymbol \delta}' \Phi(x)$ for the integration measure is equal to one. 

However, the modified BRST transformation violates the nilpotency when  $M \not= 0$:
	\begin{align}
\begin{cases}
		{\boldsymbol \delta}' {\boldsymbol \delta}' \mathscr{A}_{\mu}(x) = 0 , \\ 
		{\boldsymbol \delta}' {\boldsymbol \delta}' \mathscr{C}(x) = 0 , \\ 
		{\boldsymbol \delta}' {\boldsymbol \delta}' \bar{\mathscr{C}}(x) = i {\boldsymbol \delta}' \mathscr{N}(x)
		= i M^2 \mathscr{C}(x) \ne 0 , \\ 
		{\boldsymbol \delta}' {\boldsymbol \delta}' \mathscr{N}(x) = M^2 {\boldsymbol \delta}' \mathscr{C}(x)
	=	- M^2 \frac{g}{2} \mathscr{C}(x) \times \mathscr{C}(x) \ne 0 .
\end{cases} 
	\end{align}
The nilpotency is violated also for the modified anti-BRST transformation when  $M \not= 0$:
	\begin{align}
\begin{cases}
		\bar{{\boldsymbol \delta}}' \bar{{\boldsymbol \delta}}' \mathscr{A}_{\mu}(x) = 0 ,  \\
		\bar{{\boldsymbol \delta}}' \bar{{\boldsymbol \delta}}' \bar{\mathscr{C}}(x) = 0 , \\
		\bar{{\boldsymbol \delta}}' \bar{{\boldsymbol \delta}}' \mathscr{C}(x) = i \bar{{\boldsymbol \delta}}' \bar{\mathscr{N}}(x)
		= - i M^2 \bar{\mathscr{C}}(x) \ne 0 ,  \\
		\bar{{\boldsymbol \delta}}' \bar{{\boldsymbol \delta}}' \bar{\mathscr{N}}(x) = - M^2 \bar{{\boldsymbol \delta}}' \bar{\mathscr{C}}(x)
		= M^2 \frac{g}{2} \bar{\mathscr{C}}(x) \times \bar{\mathscr{C}}(x) \ne 0 .
\end{cases} 
	\end{align}
In the limit $M \to 0$, the modified BRST and anti-BRST transformations reduce  to the usual BRST and anti-BRST transformations and become nilpotent.

Moreover, it is checked that the modified BRST and modified anti-BRST transformations   no longer  anticommute in the $M \not=0$ case:
	\begin{align}
\begin{cases}
		\{ {\boldsymbol \delta}' , \bar{\boldsymbol \delta}' \} \mathscr{A}_{\mu}(x) = 0 , \\ 
		\{ {\boldsymbol \delta}' , \bar{\boldsymbol \delta}' \} \mathscr{C}(x) = -iM^2  \mathscr{C}(x) ,  \\ 
		\{ {\boldsymbol \delta}' , \bar{\boldsymbol \delta}' \} \bar{\mathscr{C}}(x) =  iM^2  \mathscr{\bar C}(x), \\ 
		\{ {\boldsymbol \delta}' , \bar{\boldsymbol \delta}' \} \mathscr{N}(x) =   0 .
\end{cases} 
	\end{align}	
In the limit $M \to 0$, the anticommutativity is recovered, $\{ {\boldsymbol \delta}' , \bar{\boldsymbol \delta}' \}\to 0$.

%%%%%%%%%%%%%%%%%%%%%%%%%%%%%%%%%%%%%%%%%%%%%%%%%%%%%%%%%%%%%%%%%%%%%
%%%%%%%%%%%%%%%%%%%%%%%%%%%%%%%%%%%%%%%%%%%%%%%%%%%%%%%%%%%%%%%%%%%%%
%\section{$\beta$ dependence}
%%%%%%%%%%%%%%%%%%%%%%%%%%%%%%%%%%%%%%%%%%%%%%%%%%%%%%%%%%%%%%%%%%%%%
%%%%%%%%%%%%%%%%%%%%%%%%%%%%%%%%%%%%%%%%%%%%%%%%%%%%%%%%%%%%%%%%%%%%%

Let $W$ be the generating functional of the connected Green functions defined from the vacuum functional $Z[J]$ with the source $J$ for an operator $\mathscr{O}$ as a functional of $\Phi$: 
	\begin{align}
	  e^{iW[J]} 
:=   Z[J] 
&:=   \int \mathcal{D} \mathscr{A} \mathcal{D} \mathscr{C} \mathcal{D} \bar{\mathscr{C}} \mathcal{D} \mathscr{N} 
\nonumber\\& 
\times \exp \left\{ iS^{\rm tot}_{m\rm{YM}} + i\int d^Dx J(x) \cdot \mathscr{O}(x) \right\} .
	\end{align}
Then the derivative of $W$ with  respect to $\beta$ is given by
	\begin{equation}
	\frac{\partial W[J]}{\partial \beta}
	= \frac{1}{i} \frac{\partial \ln Z[J]}{\partial \beta}
	= \frac{1}{i} Z[J]^{-1} \frac{\partial  Z[J]}{\partial \beta}
	= \Big\langle \frac{\partial S^{\rm tot}_{m\rm{YM}}}{\partial \beta} \Big\rangle_J ,
	\end{equation}
where
	\begin{equation}
	 \frac{\partial S^{\rm tot}_{m\rm{YM}}}{\partial \beta} 
= \int d^Dx \left[ i {\boldsymbol \delta}' \bar{{\boldsymbol \delta}}' \left( \frac{1}{2} i \bar{\mathscr{C}} \cdot \mathscr{C} \right) + \frac{1}{2} M^2 i \bar{\mathscr{C}} \cdot \mathscr{C} \right] .
	\end{equation}
This follows from the fact that 
$\mathscr{L}_{\rm GF+FP} + \mathscr{L}_{m}$ is written as \cite{Kondo12}
	\begin{align}
	\mathscr{L}_{\rm GF+FP} + \mathscr{L}_{m} 
=& i {\boldsymbol \delta}' \bar{{\boldsymbol \delta}}' \left(\frac{1}{2} \mathscr{A}^{\mu} \cdot \mathscr{A}_{\mu}
		+ \frac{\beta}{2} i \bar{\mathscr{C}} \cdot \mathscr{C} \right) 
		 \nonumber\\& 
+ \frac{\beta}{2} M^2 i \bar{\mathscr{C}} \cdot \mathscr{C}
+ \frac{1}{2} M^2 \mathscr{A}_{\mu} \cdot \mathscr{A}^{\mu} .
	\end{align}

If we require the modified BRST invariance of the vacuum:
	\begin{equation}
	 Q_{\rm BRST}' | 0 \rangle = 0 ,
	\end{equation}	
	we find the $\beta$ dependence of $W[J]$:
	\begin{equation}
	\frac{\partial W[J]}{\partial \beta}
	= \int d^Dx \frac{1}{2} M^2 \langle  i \bar{\mathscr{C}}(x) \cdot \mathscr{C}(x)  \rangle_J \not= 0 .
	\end{equation}
Therefore, for $M \not=0$,  $W[J]$ depends on the parameter $\beta$.	
This result should be compared with the $M=0$ case, in which $\beta$ is a gauge fixing parameter and hence  $W[J]$ should not depend on $\beta$. In the $M=0$ case, any choice of $\beta$ gives the same  $W[J]$. 
However, this is not the case for $M \not=0$.
The $\beta$ dependence of the CF model was pointed out also in \cite{Lavrov12} using different arguments.

%%%%%%%%%%%%%%%%%%%%%%%%%%%%%%%%%%%%%%%%%%%%%%%%%%%%%%%%%%%%%%%%%%%%%%%%
%%%%%%%%%%%%%%%%%%%%%%%%%%%%%%%%%%%%%%%%%%%%%%%%%%%%%%%%%%%%%%%%%%%%%%%%
\section{Defining a massive Yang-Mills field}
%%%%%%%%%%%%%%%%%%%%%%%%%%%%%%%%%%%%%%%%%%%%%%%%%%%%%%%%%%%%%%%%%%%%%%%%
%%%%%%%%%%%%%%%%%%%%%%%%%%%%%%%%%%%%%%%%%%%%%%%%%%%%%%%%%%%%%%%%%%%%%%%%

We require the following properties to construct a non-Abelian massive spin-one vector boson field $\mathscr{K}_{\mu}(x)$ in a nonperturbative way:
\renewcommand{\theenumi}{\roman{enumi}}
\renewcommand{\labelenumi}{(\theenumi)}
\begin{enumerate}
\item 
$\mathscr{K}_{\mu}$ has the  modified  BRST invariance (off mass shell): %physical field
	\begin{equation}
		{\boldsymbol \delta}' \mathscr{K}_{\mu} = 0 .
	\end{equation}

\item 
$\mathscr{K}_{\mu}$ is divergenceless (on mass shell): %3 d.o.f.
	\begin{equation}
		\partial^{\mu} \mathscr{K}_{\mu} = 0 .
	\end{equation}

\item 
$\mathscr{K}_{\mu}$ obeys the adjoint transformation under the color rotation:
	\begin{equation}
		\mathscr{K}_{\mu}(x) \to U \mathscr{K}_{\mu}(x) U^{-1} , \quad U = \exp[i \varepsilon^A Q^A] ,
	\end{equation}	
\end{enumerate}
	which has the infinitesimal version: 
	\begin{equation}
		\delta \mathscr{K}_{\mu}(x) = \varepsilon \times \mathscr{K}_{\mu}(x) . 
	\end{equation}
The field $\mathscr{K}_\mu$ is identified with the non-Abelian version of the physical massive vector field with spin one, as ensured by the above properties.
Here (i) guarantees that $\mathscr{K}_{\mu}$ belong  to the physical field creating a physical state with positive norm.
(ii)  guarantees that $\mathscr{K}_{\mu}$ have the correct degrees of freedom as a massive spin-one particle, i.e., three in the four-dimensional spacetime, i.e., two transverse and one longitudinal modes, excluding one scalar mode. 
(iii) guarantees that $\mathscr{K}_{\mu}$ obey the same transformation rule as that of the original gauge field $\mathscr{A}_{\mu}$

We observe that the total Lagrangian of the CF model is invariant under the (infinitesimal) \textbf{global gauge transformation} or \textbf{color rotation} defined by
	\begin{align}
		&\delta \Phi(x) := [\varepsilon^C i Q^C, \Phi(x)] = \varepsilon \times \Phi(x) , \nonumber\\
 & \quad {\rm for} \quad
		 \Phi=\mathscr{A}_{\mu}, \mathscr{N}, \mathscr{C} , \bar{\mathscr{C}} ,  \\
		&\delta \varphi(x) := [\varepsilon^C i Q^C, \varphi(x)] = - i \varepsilon \varphi(x) ,
	\end{align}
where $\varphi$ is a matter field, 
which is also written as
	\begin{align}
		&\delta \Phi^A(x) = f^{ABC} \varepsilon^B \Phi^C(x) , \nonumber\\
		&\delta \varphi_a(x) = - i \varepsilon^A (T^A)_a^{\ b} \varphi_b(x) = - i \varepsilon^A (T^A \varphi)_a ,
	\end{align}
where  the  conserved Noether charge $Q^A := \int d^3x \mathscr{J}^{0,A}_{\rm color}$ obtained from the color current $\mathscr{J}^0_{\rm color}$ is called the \textbf{color charge}
and is equal to the generator of the color rotation.

It has been shown \cite{Kondo12} that such a field $\mathscr{K}_{\mu}$ is obtained by a nonlinear but local transformation from the original fields  $\mathscr{A}_\mu$,  $\mathscr{C}$,  $\bar{\mathscr{C}}$ and $\mathscr{N}$ of the CF model:
\begin{align}
 \mathscr{K}_\mu :=&  \mathscr{A}_\mu - M^{-2} \partial_\mu \mathscr{N} 
- gM^{-2} \mathscr{A}_\mu \times \mathscr{N} 
\nonumber\\
&+ gM^{-2}  \partial_\mu \mathscr{C} \times i\bar{\mathscr{C}} 
+ g^2 M^{-2} (\mathscr{A}_\mu \times \mathscr{C}) \times i \bar{\mathscr{C}} .
\label{K}
\end{align}
In the Abelian limit or the lowest order of the coupling constant $g$, $\mathscr{K}_{\mu}$ reduces to the Proca field for massive vector:
	\begin{equation}
		\mathscr{K}_{\mu} \to \mathscr{A}_{\mu} - \frac{1}{M^2} \partial_{\mu} \mathscr{N} := U_{\mu} .
	\end{equation}
It should be remarked that $U_\mu$	is invariant under the Abelian version of the modified BRST, but it is not invariant under the non-Abelian modified BRST transformation.

The new field $\mathscr{K}_{\mu}$ is converted to a simple form:
	\begin{equation}
		\mathscr{K}_{\mu}(x) = \mathscr{A}_{\mu}(x) + \frac{1}{M^2} i {\boldsymbol \delta}' \bar{{\boldsymbol \delta}}' \mathscr{A}_{\mu}(x) .
		\label{K2}
	\end{equation}
It has been explicitly shown in \cite{Kondo12} that the field $\mathscr{K}_{\mu}$ defined by (\ref{K}) or (\ref{K2}) satisfies all the above properties.
The field $\mathscr{K}_{\mu}$ plays the role of the non-Abelian massive vector field
and is identified with a non-Abelian version of the spin-one massive vector field.
Equation (\ref{K}) gives a transformation from $\mathscr{A}_{\mu}, \mathscr{N}, \mathscr{C}$ and $\bar{\mathscr{C}}$ to $\mathscr{K}_{\mu}$.

As an immediate application of the above result, we can construct a mass term which is invariant simultaneously under the modified BRST transformation,  Lorentz transformation and  color rotation:
	\begin{equation}
		\frac{1}{2} M^2 \mathscr{K}_{\mu}(x) \cdot \mathscr{K}^{\mu}(x) .
	\end{equation}
This can be useful as a regularization scheme for avoiding infrared divergences in non-Abelian gauge theories. 
Moreover, we can obtain a dimension-two condensate which is modified  BRST invariant, Lorentz invariant, and color-singlet: 
	\begin{equation}
		\langle \mathscr{K}_{\mu}(x) \cdot \mathscr{K}^{\mu}(x) \rangle .
	\end{equation}
	This dimension-two condensate is off-shell (modified) BRST invariant and  should be compared with the dimension-two condensate proposed in \cite{Kondo01,KMSI02}:
	\begin{equation}
		\Big\langle \frac12 \mathscr{A}_{\mu}(x) \cdot \mathscr{A}^{\mu}(x) + \beta  \mathscr{C}(x) \cdot \mathscr{\bar C}(x) \Big\rangle ,
	\end{equation}
which is only on-shell BRST invariant.

The original CF Lagrangian $\mathscr{L}_{\rm mYM}^{\rm tot}[\mathscr{A}_\mu,\mathscr{C},\bar{\mathscr{C}},\mathscr{N}]$ is written in terms of $\mathscr{A}_\mu, \mathscr{C}, \bar{\mathscr{C}}$ and $\mathscr{N}$.
The new theory is specified by $\mathscr{L}_{\rm mYM}^{\rm tot}[\mathscr{K}_\mu,\mathscr{C},\bar{\mathscr{C}},\mathscr{N}]$ written in terms of $\mathscr{K}_\mu, \mathscr{C}, \bar{\mathscr{C}}$ and $\mathscr{N}$ with the symmetry:
\begin{align}   
\begin{cases}
			{\boldsymbol \delta}^\prime \mathscr{K}_{\mu}(x) =  0
\\
			{\boldsymbol \delta}^\prime \mathscr{C}(x) %= ig \{C(x), C(x) \} (
=  -\frac{g}{2} \mathscr{C}(x) \times \mathscr{C}(x)  
\\
			{\boldsymbol \delta}^\prime \bar{\mathscr{C}}(x) =  i \mathscr{N}(x) 
\\
			{\boldsymbol \delta}^\prime \mathscr{N}(x) =  M^2 \mathscr{C}(x) 
\end{cases} 
 .
 \label{Sym2}
\end{align}

%%%%%%%%%%%%%%%%%%%%%%%%%%%%%%%%%%%%%%%%%%%%%%%%%%%%%%%%%%%%%%%%%%%%%%%%
%%%%%%%%%%%%%%%%%%%%%%%%%%%%%%%%%%%%%%%%%%%%%%%%%%%%%%%%%%%%%%%%%%%%%%%%
\section{Perturbative framework of the massive Yang-Mills theory}
%%%%%%%%%%%%%%%%%%%%%%%%%%%%%%%%%%%%%%%%%%%%%%%%%%%%%%%%%%%%%%%%%%%%%%%%
%%%%%%%%%%%%%%%%%%%%%%%%%%%%%%%%%%%%%%%%%%%%%%%%%%%%%%%%%%%%%%%%%%%%%%%%

Equation (\ref{K}) gives a transformation from $\mathscr{A}_\mu, \mathscr{C}, \bar{\mathscr{C}}$ and $\mathscr{N}$ to $\mathscr{K}_\mu$.
In order to write explicitly  the non-Abelian massive Yang-Mills theory without the Higgs field, 
we rewrite the original Lagrangian written in terms of $\mathscr{A}_{\mu}, \mathscr{N}, \mathscr{C}$ and $\bar{\mathscr{C}}$
into the new Lagrangian written in terms of $\mathscr{K}_{\mu}, \mathscr{N}, \mathscr{C}$ and $\bar{\mathscr{C}}$.
For this purpose, we need the inverse transformation of (\ref{K}), namely $\mathscr{A}_{\mu}$ as a function of $\mathscr{K}_{\mu}, \mathscr{N}, \mathscr{C}$ and $\bar{\mathscr{C}}$.
But the inverse transformation cannot be given in a closed form,
since  (\ref{K}) is a nonlinear transformation.
In order to perform the perturbative calculation, it is sufficient to know the order by order expression of the inverse transformation.
By using (\ref{K}) in the form:
\begin{align}
 \mathscr{A}_\mu =&  \mathscr{K}_\mu + M^{-2} \partial_\mu \mathscr{N} 
+ gM^{-2} \mathscr{A}_\mu \times \mathscr{N} 
\nonumber\\
&- gM^{-2}  \partial_\mu \mathscr{C} \times i\bar{\mathscr{C}} 
- g^2 M^{-2} (\mathscr{A}_\mu \times \mathscr{C}) \times i \bar{\mathscr{C}} ,
	\label{A_mu}
\end{align}
we find that  the right-hand side contains the order $g^0, g^1$ and $g^2$ terms.
By iterative procedures, we obtain
\begin{align}
 \mathscr{A}_\mu =& \mathscr{K}_\mu + M^{-2} \partial_\mu \mathscr{N} 
- gM^{-2} i\bar{\mathscr{C}} \times  \partial_\mu \mathscr{C}  
\nonumber\\
&+ gM^{-2} \mathscr{K}_\mu \times \mathscr{N} 
+ gM^{-4} \partial_\mu \mathscr{N} \times \mathscr{N} + O(g^2) .
		\label{A_mu-expan}
\end{align}
The propagator of $\mathscr{K}_{\mu}$ is obtained from the order $g^0$ terms, i.e.,
$\mathscr{A}_{\mu} = \mathscr{K}_{\mu} + \frac{1}{M^2} \partial_{\mu} \mathscr{N}$, as in the Abelian case, and hence the propagator is not modified from the Proca case. 
However, the vertex functions among $\mathscr{K}_\mu, \mathscr{C}, \bar{\mathscr{C}}$ and $\mathscr{N}$  in the massive theory are modified by the order $g^1$ terms from those of  $\mathscr{A}_\mu, \mathscr{C}, \bar{\mathscr{C}}$ and $\mathscr{N}$ in the original theory.
This fact has been overlooked in all the preceding studies.  
The preceding works \cite{DV70,SF70,Boulware70,CF76,CF76b,Ojima82,BSNW96} are based on the observation that the vertices in the massive case are the same as the massless case. 
However, this observation is correct only if the relationship between the original field $\mathscr{A}_\mu$ to the massive field $\mathscr{K}_\mu$ is linear as in the Abelian case.  
This is not the case in the non-Abelian case, as our construction of the massive field $\mathscr{K}_\mu$ clearly shows. 
The vertices in the massive theory are modified in terms of the vector field $\mathscr{K}_\mu$ in addition to the FP ghost, the FP antighost and the NL field.
%, so that the physical unitarity holds even in the massive theory. 
%This is the main reason why the physical unitarity was seemed to be violated so far. 
For the correct identification of the massive vector field $\mathscr{K}_\mu$, one needs the FP ghost, the FP antighost and the NL field, in addition to the original Yang-Mills field $\mathscr{A}_\mu$.
This could be a loophole of avoiding the results of the preceding analyses for the violation of physical unitarity. 
It is checked if this is true or not. 
%This is a main reason why the physical unitarity looked violated so far.

To check the physical unitarity in the nontrivial lowest order in the perturbation with respect to the coupling constant $g$,
the original Lagrangian $\mathscr{L}^{\rm{tot}}_{m\rm{YM}}$ is rewritten order by order of the coupling constant $g$ as follows.
	\begin{equation}
		\mathscr{L}_{\rm CF} = \mathscr{L}_0 + \mathscr{L}_1 + O(g^2) ,
	\end{equation}
	\begin{align}
		\mathscr{L}_0 =& - \frac{1}{4} (\partial_{\mu} \mathscr{K}_{\nu} - \partial_{\nu} \mathscr{K}_{\mu} )^2
		+ \frac{1}{2} M^2 \mathscr{K}_{\mu} \cdot \mathscr{K}^{\mu} \nonumber\\
		& - \frac{1}{2M^2} (\partial^{\mu} \mathscr{N})^2 + \frac{\beta}{2} \mathscr{N} \cdot \mathscr{N} \nonumber\\
		& - i \partial^{\mu} \bar{\mathscr{C}} \cdot \partial_{\mu} \mathscr{C} 
		+\beta M^2 i \bar{\mathscr{C}} \cdot \mathscr{C} ,
		\label{L0}
	\end{align}
	\begin{align}
		\mathscr{L}_1
		=& - \frac{g}{2} (\partial_{\mu} \mathscr{K}_{\nu} - \partial_{\nu} \mathscr{K}_{\mu}) \cdot (\mathscr{K}^{\mu} \times \mathscr{K}^{\nu}) \nonumber\\
		&+ \frac{g}{2M^4} (\partial_{\mu} \mathscr{K}_{\nu} - \partial_{\nu} \mathscr{K}_{\mu})
		\cdot (\partial^{\mu} \mathscr{N} \times \partial^{\nu} \mathscr{N})
 \nonumber\\&
		  - \frac{g}{M^2} \mathscr{K}_{\mu} \cdot (\mathscr{N} \times \partial^{\mu} \mathscr{N})  \nonumber\\
		&+ \frac{g}{2M^2} (\partial_{\mu} \mathscr{K}_{\nu} - \partial_{\nu} \mathscr{K}_{\mu})
		\cdot (i \partial^{\mu} \bar{\mathscr{C}} \times \partial^{\nu} \mathscr{C} \nonumber\\
		&- i \partial^{\nu} \bar{\mathscr{C}} \times \partial^{\mu} \mathscr{C}) \nonumber\\
		&- g \mathscr{K}_{\mu} \cdot (i \bar{\mathscr{C}} \times \partial^{\mu} \mathscr{C})
%		\nonumber\\&
		 + g \mathscr{K}_{\mu} \cdot (i \partial^{\mu} \bar{\mathscr{C}} \times \mathscr{C}) \nonumber\\
		&+ \frac{g}{M^2} \partial^{\mu} \mathscr{N} \cdot (i \partial_{\mu} \bar{\mathscr{C}} \times \mathscr{C})
		- g \frac{\beta}{2} \mathscr{N} \cdot (i \bar{\mathscr{C}} \times \mathscr{C}) .
		\label{L1}
	\end{align}
The Feynman rules are given up to three-point vertices of $O(g)$ as follows.

\underline{Propagators} (Fig.~\ref{fig:feynman-rule-mYM-prop}) \\
\renewcommand{\theenumi}{\alph{enumi}}
\renewcommand{\labelenumi}{(\theenumi)}
\begin{enumerate}
\item Massive vector propagator $\langle  {\mathscr{K}}^A_{\mu}(k)  {\mathscr{K}}^B_{\nu}(-k) \rangle$
	\begin{equation}
%		\langle  {\mathscr{K}}^A_{\mu}(k)  {\mathscr{K}}^B_{\nu}(-k) \rangle =
 - \frac{i}{k^2- M^2 + i \varepsilon} \left( g_{\mu \nu} - \frac{k_{\mu}k_{\nu}}{M^2}\right) \delta^{AB} 
		%=  \Delta^{AB}_{\mu\nu}(k) .
		\label{KK-propagator}
	\end{equation}

\item FP ghost propagator $\langle  {\mathscr{C}}^A(k)  {\bar{\mathscr{C}}}^B(-k) \rangle$
	\begin{equation}
%	\langle  {\mathscr{C}}^A(k)  {\bar{\mathscr{C}}}^B(-k) \rangle 	 =
 - \frac{i}{k^2- \beta M^2 + i \varepsilon} \delta^{AB}
		= \Delta^{AB}(k) .
	\end{equation}

\item NL (auxiliary) field propagator $\langle  {\mathscr{N}}^A(k)  {\mathscr{N}}^B(-k) \rangle$
	\begin{equation}
%		\langle  {\mathscr{N}}^A(k)  {\mathscr{N}}^B(-k) \rangle =  
 - \frac{iM^2}{k^2- \beta M^2 + i \varepsilon} \delta^{AB}    .
	\end{equation}

%%%%%%%%%%%%%%%%%%%%%%%%%%%%%%%%%%%%%%%%%%%%%%%%%%%%%%%%%%%%%%%%%%%
\begin{figure}[tb]% 
%\vspace{30mm}
\begin{center}
\includegraphics[scale=0.5]{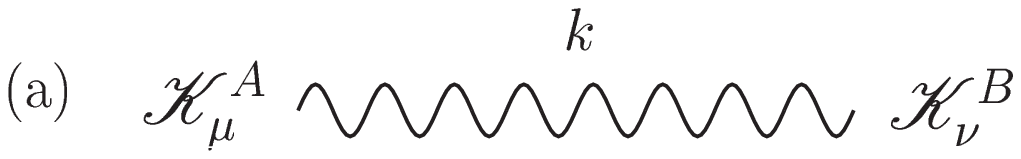}
\includegraphics[scale=0.5]{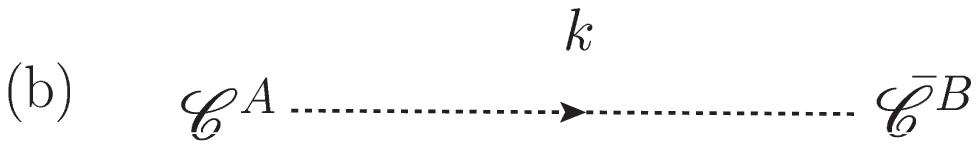}
\includegraphics[scale=0.5]{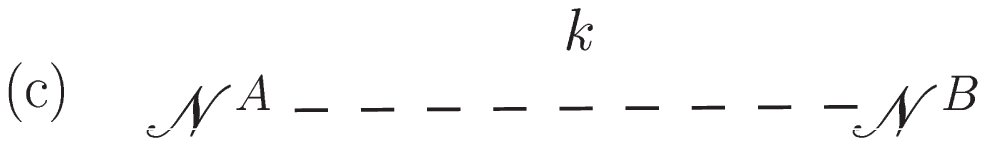}
\end{center}
%\capwidth90mm
\caption{
Feynman rules for propagators: 
(a)~massive vector propagator,   (b)~FP ghost propagator,   (c)~NL field propagator.
}
\label{fig:feynman-rule-mYM-prop}
\end{figure}
%%%%%%%%%%%%%%%%%%%%%%%%%%%%%%%%%%%%%%%%%%%%%%%%%%%%%%%%%%%%%%%%%%%

%%%%%%%%%%%%%%%%%%%%%%%%%%%%%%%%%%%%%%%%%%%%%%%%%%%%%%%%%%%%%%%%%%%
\begin{figure}[tb]% 
%\vspace{30mm}
\begin{center}
\includegraphics[scale=0.5]{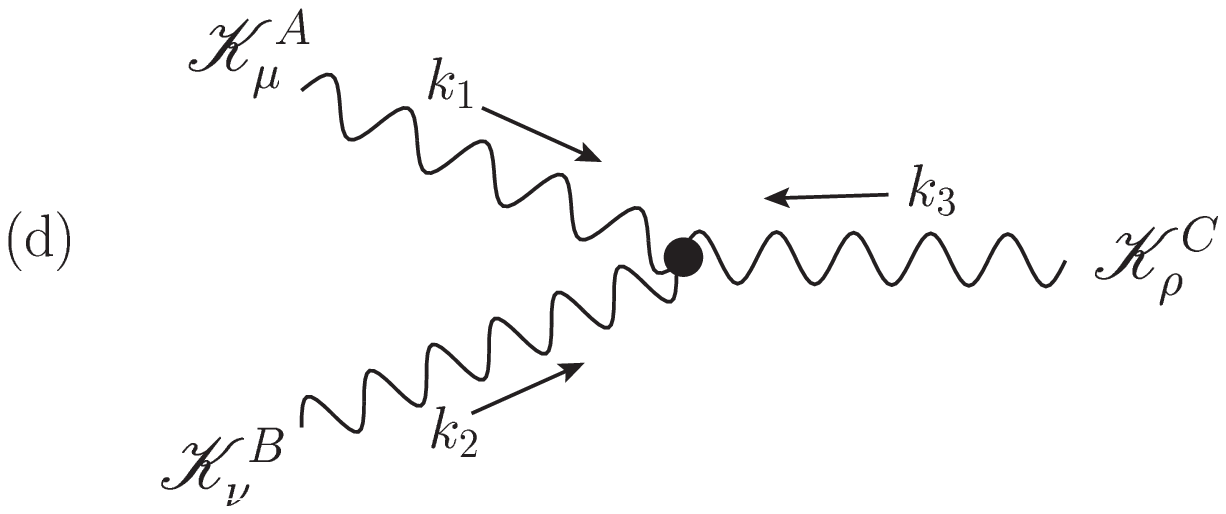}
\includegraphics[scale=0.5]{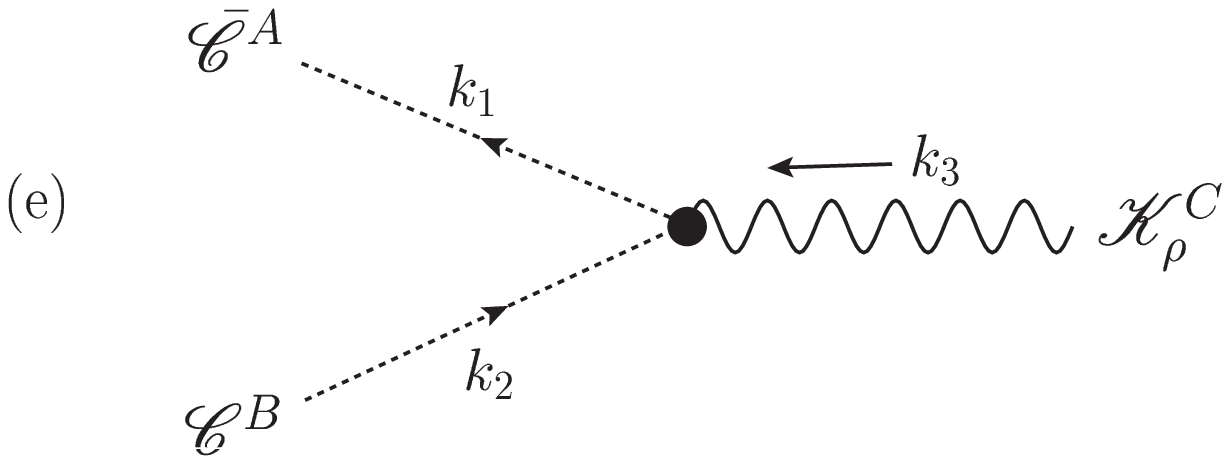}
\includegraphics[scale=0.5]{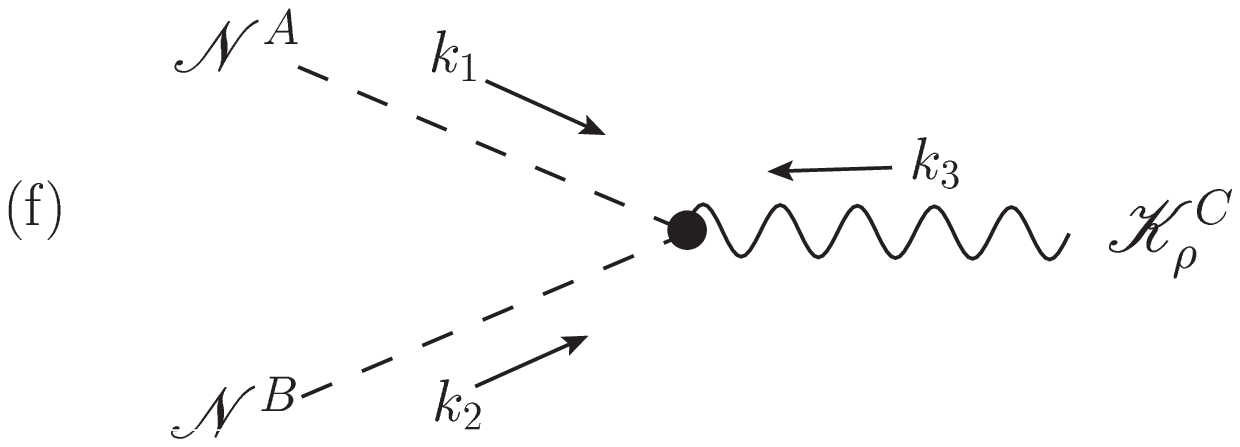}
\includegraphics[scale=0.5]{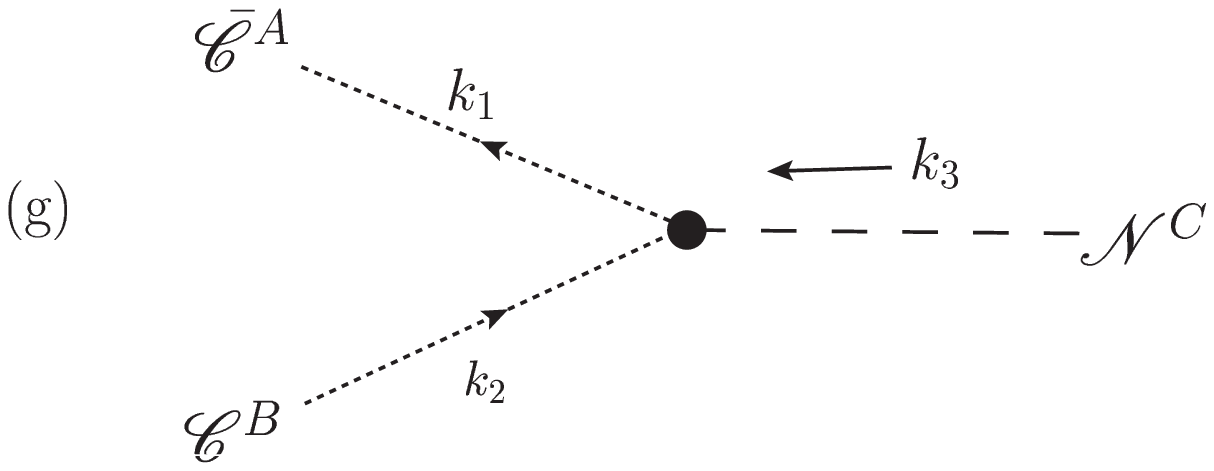}
\end{center}
%\capwidth90mm
\caption{
Feynman rules for three-point vertex functions: 
(d) $\mathscr{K} \mathscr{K} \mathscr{K}$ vertex,  (e) $\mathscr{K} \mathscr{C} \bar{\mathscr{C}}$ vertex,   (f)  $\mathscr{K} \mathscr{N} \mathscr{N}$ vertex, (g) $\mathscr{N} \mathscr{C} \bar{\mathscr{C}}$ vertex.
}
\label{fig:feynman-rule-mYM-vertex}
\end{figure}
%%%%%%%%%%%%%%%%%%%%%%%%%%%%%%%%%%%%%%%%%%%%%%%%%%%%%%%%%%%%%%%%%%%

\underline{Three-point vertices} (Fig.~\ref{fig:feynman-rule-mYM-vertex}) \\
\item $\mathscr{K} \mathscr{K} \mathscr{K}$ vertex $\langle \mathscr{K}^A_{\mu}(k_1) \mathscr{K}^B_{\nu}(k_2) \mathscr{K}^C_{\rho}(k_3) \rangle$ %Three-vectors  ($\mathscr{K} \mathscr{K} \mathscr{K}$)
	\begin{align}
%\langle \mathscr{K}^A_{\mu}(k_1) \mathscr{K}^B_{\nu}(k_2) \mathscr{K}^C_{\rho}(k_3) \rangle %\nonumber\\
%		=& - i g f^{ABC} [(k_1 - k_2)_{\rho} g_{\mu\nu} + (k_2-k_3)_{\mu}g_{\nu\rho} + (k_3-k_1)_{\nu}g_{\rho\mu}]		=
  -ig f^{ABC} V_{\mu\nu\rho}(k_1, k_2, k_3) ,
	\end{align}
with
	\begin{align}
		V_{\mu\nu\rho}(k_1, k_2, k_3) :=& (k_1-k_2)_{\rho}g_{\mu\nu} + (k_2-k_3)_{\mu}g_{\nu\rho} \nonumber\\
		&+ (k_3-k_1)_{\nu}g_{\rho\mu} .
	\end{align}

\item  $\mathscr{K} \mathscr{C} \bar{\mathscr{C}}$ vertex $\langle \bar{\mathscr{C}}^A(k_1) \mathscr{C}^B(k_2) \mathscr{K}^C_{\rho}(k_3) \rangle$ 
%One vector, one ghost and one anti-ghost ($\mathscr{K} \mathscr{C} \bar{\mathscr{C}}$)
	\begin{align}
%	&\langle \bar{\mathscr{C}}^A(k_1) \mathscr{C}^B(k_2) \mathscr{K}^C_{\rho}(k_3) \rangle \nonumber\\		=& 
&-ig f^{ABC} \{ M^{-2} [k_{1\rho} (k_2 \cdot k_3) - k_{2\rho} (k_1 \cdot k_3)] 
\nonumber\\&
+ k_{1\rho} - k_{2\rho} \} .
	\end{align}

\item $\mathscr{K} \mathscr{N} \mathscr{N}$ vertex $\langle \mathscr{N}^A(k_1) \mathscr{N}^B(k_2) \mathscr{K}^C_{\rho}(k_3) \rangle$ 
%One vector and two NL fields %($\mathscr{K} \mathscr{N} \mathscr{N}$)
	\begin{align}
%		&\langle \mathscr{N}^A(k_1) \mathscr{N}^B(k_2) \mathscr{K}^C_{\rho}(k_3) \rangle \nonumber\\		=& 
& i gM^{-2} f^{ABC} \{ M^{-2} [k_{1\rho} (k_2 \cdot k_3) - k_{2\rho} (k_1 \cdot k_3)] \nonumber\\&
+ k_{1\rho} - k_{2\rho} \} .
	\end{align}

\item $\mathscr{N} \mathscr{C} \bar{\mathscr{C}}$ vertex $\langle \bar{\mathscr{C}}^A(k_1) \mathscr{C}^B(k_2) \mathscr{N}^C(k_3) \rangle$ %One NL field, one ghost and one anti-ghost  ($\mathscr{N} \mathscr{C} \bar{\mathscr{C}}$)
	\begin{align}
%	\langle \bar{\mathscr{C}}^A(k_1) \mathscr{C}^B(k_2) \mathscr{N}^C(k_3) \rangle %\nonumber\\		=  
ig f_{ABC} [M^{-2} k_1 \cdot k_3 -  \beta/2] . 
	\end{align}

\end{enumerate}

%%%%%%%%%%%%%%%%%%%%%%%%%%%%%%%%%%%%%%%%%%%%%%%%%%%%%%%%%%%%%%%%%%%
\begin{figure}[tb]% 
%\vspace{30mm}
\begin{center}
\includegraphics[scale=0.5]{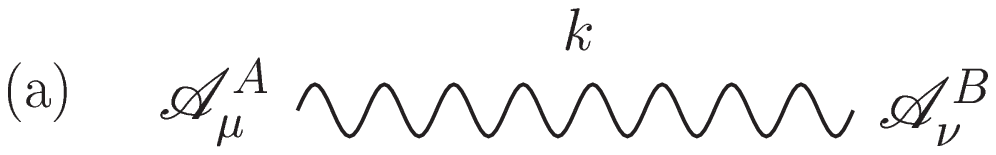}
\includegraphics[scale=0.5]{fig-ep196/FP_ghost_propagator.eps}
\includegraphics[scale=0.5]{fig-ep196/NL_field_propagator.eps}
\includegraphics[scale=0.5]{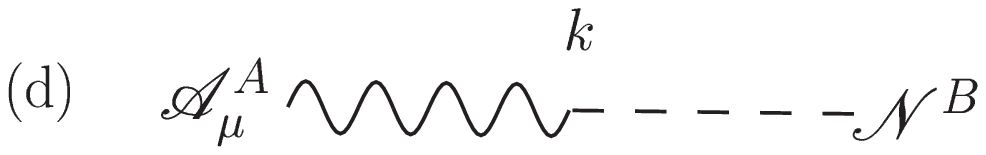}
\end{center}
%\capwidth90mm
\caption{
Feynman rules for propagators:  (a)~massive vector propagator,  (b)~FP ghost propagator,  (c)~NL field propagator, (d)~cross propagator.
}
\label{fig:feynman-rule-mYM-prop-0}
\end{figure}
%%%%%%%%%%%%%%%%%%%%%%%%%%%%%%%%%%%%%%%%%%%%%%%%%%%%%%%%%%%%%%%%%%%

%%%%%%%%%%%%%%%%%%%%%%%%%%%%%%%%%%%%%%%%%%%%%%%%%%%%%%%%%%%%%%%%%%%
\begin{figure}[tb]% 
%\vspace{30mm}
\begin{center}
\includegraphics[scale=0.5]{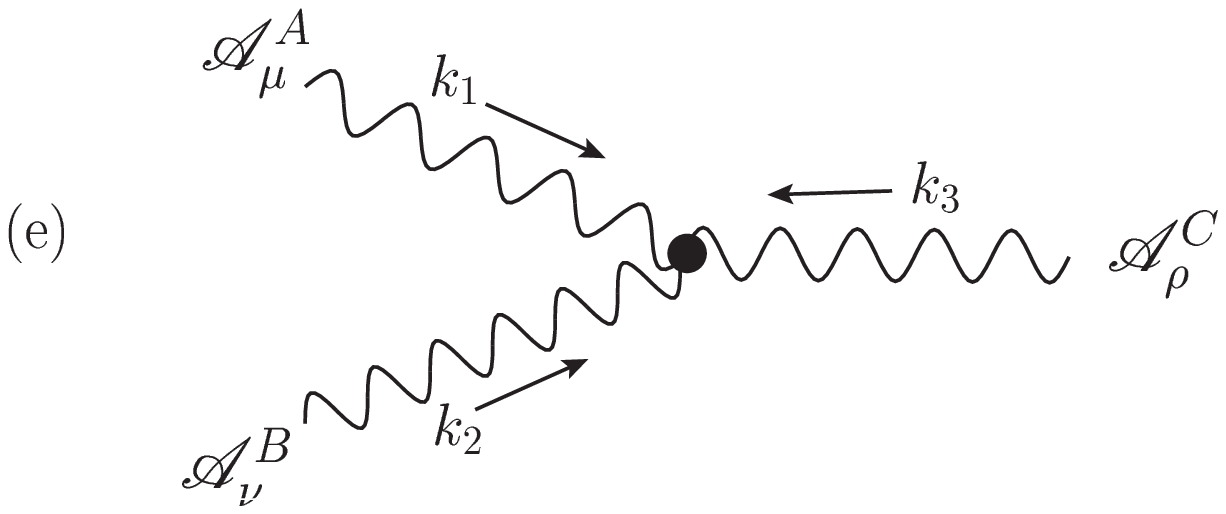}
\includegraphics[scale=0.5]{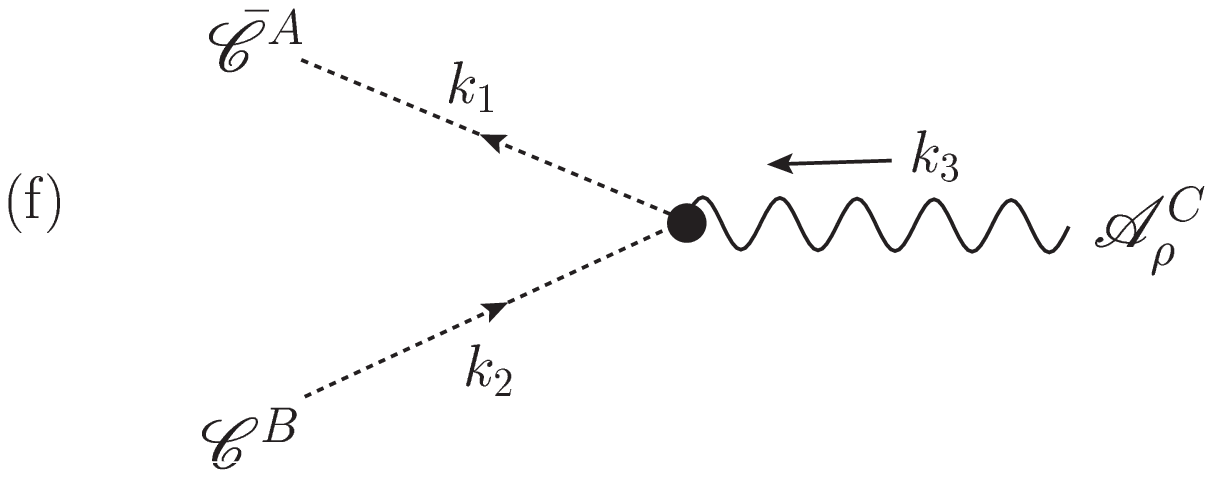}
\end{center}
%\capwidth90mm
\caption{
Feynman rules for three-point vertex functions:   (e) $\mathscr{A} \mathscr{A} \mathscr{A}$ vertex, (f) $\mathscr{A} \mathscr{C} \bar{\mathscr{C}}$ vertex.
}
\label{fig:feynman-rule-mYM-vertex-0}
\end{figure}
%%%%%%%%%%%%%%%%%%%%%%%%%%%%%%%%%%%%%%%%%%%%%%%%%%%%%%%%%%%%%%%%%%%

These rules should be compared with the following Feynman rules for the original Lagrangian. \\
\underline{Propagators}  (Fig.~\ref{fig:feynman-rule-mYM-prop-0}) \\
\renewcommand{\theenumi}{\alph{enumi}}
\renewcommand{\labelenumi}{(\theenumi)}
\begin{enumerate}
\item Vector propagator $\langle  {\mathscr{A}}^A_{\mu}(k)  {\mathscr{A}}^B_{\nu}(-k) \rangle$
	\begin{align}
%	\langle  {\mathscr{A}}^A_{\mu}(k)  {\mathscr{A}}^B_{\nu}(-k) \rangle =&
 &-   \frac{i}{k^2- M^2 + i \varepsilon} \left( g_{\mu \nu} - \frac{k_{\mu}k_{\nu}}{k^2+ i\varepsilon}\right) \delta^{AB} \nonumber\\
		&- \frac{i\beta}{k^2 - \beta M^2 + i \varepsilon} \frac{k_{\mu}k_{\nu}}{k^2+ i \varepsilon}   \delta^{AB}
		=  D^{AB}_{\mu\nu}(k) .
		\label{AA-propagator}
	\end{align}

\item FP ghost propagator $\langle  {\mathscr{C}}^A(k)  {\bar{\mathscr{C}}}^B(-k) \rangle$
	\begin{equation}
%	 \langle  {\mathscr{C}}^A(k)  {\bar{\mathscr{C}}}^B(-k) \rangle =
	- \frac{i}{k^2- \beta M^2+ i \varepsilon} \delta^{AB} .
	\end{equation}

\item NL field propagator $\langle  {\mathscr{N}}^A(k)  {\mathscr{N}}^B(-k) \rangle$
	\begin{equation}
%	 \langle  {\mathscr{N}}^A(k)  {\mathscr{N}}^B(-k) \rangle	=
 \frac{iM^2}{k^2- \beta M^2 + i \varepsilon} \delta^{AB}  .
	\end{equation}

\item The cross propagator $\langle  {\mathscr{A}}^A_{\mu}(k)  {\mathscr{N}}^B(-k) \rangle$
	\begin{equation}
%	\langle  {\mathscr{A}}^A_{\mu}(k)  {\mathscr{N}}^B(-k) \rangle   =
	\frac{- k_{\mu}}{k^2- \beta M^2 + i \varepsilon} \delta^{AB}  .
	\end{equation}
	
\underline{Three-point vertices}  (Fig.~\ref{fig:feynman-rule-mYM-vertex-0}) \\
\item  $\mathscr{A} \mathscr{A} \mathscr{A}$ vertex $\langle \mathscr{A}_\mu^A(k_1) \mathscr{A}_\nu^B(k_2) \mathscr{A}_\rho^C(k_3) \rangle$ 
%Three-vectors ($\mathscr{A} \mathscr{A} \mathscr{A}$) 
	\begin{equation}
%\langle \mathscr{A}_\mu^A(k_1) \mathscr{A}_\nu^B(k_2) \mathscr{A}_\rho^C(k_3) \rangle = 
 -ig f^{ABC} V_{\mu\nu\rho}(k_1, k_2, k_3) ,
%= (\ref{V1'})
	\end{equation}
%\begin{equation}
%  ig f_{ABC} V_{\mu\nu\rho}(k_1,k_2,k_3), 
%  \label{V1}
%\end{equation}

\item $\mathscr{A} \mathscr{C} \bar{\mathscr{C}}$ vertex  $\langle \bar{\mathscr{C}}^A(k_1) \mathscr{C}^B(k_2) \mathscr{A}_\rho^C(k_3) \rangle$ 
%One vector, one ghost and one anti-ghost ($\mathscr{A} \mathscr{C} \bar{\mathscr{C}}$)
	\begin{equation}
%\langle \bar{\mathscr{C}}^A(k_1) \mathscr{C}^B(k_2) \mathscr{A}_\rho^C(k_3) \rangle =   
ig f_{ABC} k_{1\rho} . 
  \label{V2}
\end{equation}

\end{enumerate}

From the relation,
	\begin{equation}
		\mathscr{A}_{\mu} = \mathscr{K}_{\mu} + \frac{1}{M^2} \partial_{\mu} \mathscr{N} + O(g) ,
	\end{equation}
we find  
	\begin{align}
	&	\langle \mathscr{A}_{\mu}(x) \mathscr{A}_{\nu}(y) \rangle \nonumber\\
		=& \langle (\mathscr{K}_{\mu}(x) + M^{-2} \partial_{\mu} \mathscr{N}(x)) (\mathscr{K}_{\nu}(y) + M^{-2} \partial_{\nu} \mathscr{N}(y)) \rangle \nonumber\\
		=& \langle \mathscr{K}_{\mu}(x) \mathscr{K}_{\nu}(y) \rangle + M^{-2} \langle \mathscr{K}_{\mu}(x) \partial_{\nu} \mathscr{N}(y) \rangle \nonumber\\
		&+ M^{-2} \langle \partial_{\mu} \mathscr{N}(x) \mathscr{K}_{\nu}(y) \rangle 
+ M^{-4} \langle \partial_{\mu} \mathscr{N}(x) \partial_{\nu} \mathscr{N}(y) \rangle ,
	\end{align}
which has the Fourier transform:	
	\begin{align}
	&	\langle \tilde{\mathscr{A}}^A_{\mu}(k) \tilde{\mathscr{A}}^B_{\nu}(-k) \rangle
\nonumber\\
		=& \langle \tilde{\mathscr{K}}^A_{\mu}(k) \tilde{\mathscr{K}}^B_{\nu}(-k) \rangle
		+ ik_{\nu} M^{-2} \langle \mathscr{K}_{\mu}^A(-k) \mathscr{N}^B(k) \rangle 
\nonumber\\
		&- ik_{\mu} M^{-2} \langle \mathscr{N}^A(k) \mathscr{K}_{\nu}^B(-k) \rangle
\nonumber\\
		&+ M^{-4} k_{\mu} k_{\nu} \langle \mathscr{N}^A(k) \mathscr{N}^B(-k) \rangle 
\nonumber\\
		=& i\delta^{AB} \left[ - \frac{g_{\mu \nu} - \frac{k_{\mu} k_{\nu}}{M^2}}{k^2-M^2+i\epsilon} -   \frac{\frac{k_{\mu}k_{\nu}}{M^2}}{k^2-\beta M^2+i\epsilon} \right]
 , 
 \label{AA-propagator2}
	\end{align}
where we have used (\ref{KK-propagator}) and (\ref{AA-propagator}) and  the fact that there are no mixing propagators:
	\begin{align}
 \langle \mathscr{K}_{\mu}^A(-k) \mathscr{N}^B(k) \rangle 
= 0 = \langle \mathscr{N}^A(k) \mathscr{K}_{\nu}^B(-k) \rangle .
	\end{align}
Thus the original gluon two-point function or propagator is decomposed into the spin-one and spin-sero parts.
	
Another expression for the propagator is the manifestly (power-counting) renormalizable form:
	\begin{align}
	&	\langle \tilde{\mathscr{A}}^A_{\mu}(k) \tilde{\mathscr{A}}^B_{\nu}(-k) \rangle
%	\nonumber\\
%		=&   - i\delta^{AB} \frac{g_{\mu \nu} - \frac{k_{\mu} k_{\nu}}{k^2+i\epsilon} + \left(\frac{1}{k^2+i\epsilon}-\frac{1}{M^2}\right)k_{\mu}k_{\nu}}{k^2-M^2+i\epsilon}
%		- i\delta^{AB} \frac{\frac{k_{\mu}k_{\nu}}{M^2}}{k^2-\beta M^2+i\epsilon} 
% \nonumber\\
%		=& - i\delta^{AB} \frac{g_{\mu \nu} - \frac{k_{\mu} k_{\nu}}{k^2+i\epsilon}}{k^2-M^2+i\epsilon} \nonumber\\&
%- i\delta^{AB}\frac{M^2 - (k^2+i\epsilon)}{M^2(k^2+i\epsilon)} \frac{k_{\mu}k_{\nu}}{k^2-M^2+i\epsilon}
%		- i\delta^{AB} \frac{\frac{k_{\mu}k_{\nu}}{M^2}}{k^2-\beta M^2+i\epsilon}  
\nonumber\\
%		=& - \frac{g_{\mu \nu} - \frac{k_{\mu} k_{\nu}}{k^2}}{k^2-M^2}  + \frac{k_{\mu}k_{\nu}}{M^2} \left(\frac{1}{k^2}-\frac{1}{k^2 - \beta M^2}\right) 
%\nonumber\\
%		=& - \frac{g_{\mu \nu} - \frac{k_{\mu} k_{\nu}}{k^2}}{k^2-M^2} + \frac{k_{\mu} k_{\nu}}{M^2} \frac{-\beta M^2}{k^2(k^2-\beta M^2)} 
%\nonumber\\
		=&  i\delta^{AB} \left[ - \frac{g_{\mu \nu} - \frac{k_{\mu} k_{\nu}}{k^2+i\epsilon}}{k^2-M^2+i\epsilon} -    \frac{\beta \frac{k_{\mu}k_{\nu}}{k^2+i\epsilon}}{k^2-\beta M^2+i\epsilon} \right] .
 \label{AA-propagator3}
	\end{align}
%\footnote{
%$\beta=1$
%	\begin{equation}
%		\Rightarrow - \left[ \frac{  g_{\mu \nu} - \frac{k_{\mu} k_{\nu}}{k^2}  }{k^2-M^2}  + \frac{\frac{k_{\mu}k_{\nu}}{k^2}}{k^2-M^2} \right]
%		= - \frac{g_{\mu \nu} - \frac{k_{\mu}k_{\nu}}{M^2}}{k^2 - M^2} - \frac{\frac{k_{\mu} k_{\nu}}{M^2}}{k^2 - M^2} ,
%	\end{equation}
%	\begin{equation}
%		- \frac{g_{\mu \nu}}{k^2 - M^2}
%		= - \frac{g_{\mu \nu} - \frac{k_{\mu}k_{\nu}}{M^2}}{k^2 - M^2} - \frac{\frac{k_{\mu} k_{\nu}}{M^2}}{k^2 - M^2} .
%	\end{equation}
%}	

For higher orders, see Appendix B.

%%%%%%%%%%%%%%%%%%%%%%%%%%%%%%%%%%%%%%%%%%%%%%%%%%%%%%%%%%%%
%%%%%%%%%%%%%%%%%%%%%%%%%%%%%%%%%%%%%%%%%%%%%%%%%%%%%%%%%%%%
\section{Physical unitarity of massless Yang-Mills  theory}
%%%%%%%%%%%%%%%%%%%%%%%%%%%%%%%%%%%%%%%%%%%%%%%%%%%%%%%%%%%%
%%%%%%%%%%%%%%%%%%%%%%%%%%%%%%%%%%%%%%%%%%%%%%%%%%%%%%%%%%%%

The $S$ matrix or the scattering operator $S$ is unitary:
	\begin{equation}
		\textbf{1} = S^{\dagger}S = SS^{\dagger} .
	\end{equation}
This means that for any (initial) state $\Psi_b \in \mathcal{V}$ and any (final) state $\Psi_a \in \mathcal{V}$,
the following relation holds:
	\begin{align}
		\langle \Psi_b | \Psi_a \rangle =& \langle \Psi_b |S^{\dagger}S| \Psi_a \rangle %\nonumber\\&
		=  \sum_{\Phi_n \in \mathcal{V}} \langle \Psi_b |S^{\dagger}| \Phi_n \rangle \langle \Phi_n| S | \Psi_a \rangle ,
		\label{u-rel-1}
	\end{align}
which is obtained by inserting  the complete set of states $\{\Phi_n\}$ in the total state space $\mathcal{V}$:
$
		\textbf{1} = \sum_{\Phi_n \in \mathcal{V}} | \Phi_n \rangle \langle \Phi_n| .
$

On the other hand, the \textit{physical unitarity} of the $S$ matrix means that the $S$ matrix is unitary on the physical subspace $\mathcal{V}_{\rm phys}$: for any   physical state $\Psi_a, \Psi_b \in \mathcal{V}_{\rm phys}$,
	\begin{align}
%	& \text{for any state} \  \Psi_a, \Psi_b \in \mathcal{V}_{\rm phys}, 		\nonumber\\&
		\langle \Psi_b | \Psi_a \rangle
		= \sum_{\Phi_n \in \mathcal{V}_{\rm phys}} \langle \Psi_b |S^{\dagger}| \Phi_n \rangle \langle \Phi_n| S | \Psi_a \rangle ,
	\end{align}
where 	the physical subspace $\mathcal{V}_{\rm phys}$ is defined by
	\begin{equation}
\mathcal{V}_{\rm phys} := \{    |\rm{phys} \rangle \in \mathcal{V}  ; \langle  \rm{phys}  |  \rm{phys} \rangle \ge 0  \} \subset \mathcal{V} .
	\end{equation}
%where we have used the complete set of $\mathcal{V}_{\rm phys}$.
%$\mathscr{V}_{\rm phys} := \{    |\rm{phys} \rangle \in \mathscr{V}  ; \langle  \rm{phys}  |  \rm{phys} \rangle \ge 0  \} \subset \mathscr{V}$ 

The  unitarity of the \textbf{$S$ matrix} is rewritten in terms of 
  the \textbf{scattering amplitude} defined  by 
	\begin{equation}
		S = {\boldsymbol 1} + i T , 
%\quad {\rm or} \quad S_{ab} = \delta_{ab} + i (2\pi)^D \delta^D(p_a-p_b) T_{ab} , 
	\end{equation} 
into the relation:
	\begin{equation}
		- i (T - T^{\dagger}) = T T^{\dagger} . 
%\quad {\rm or} \quad 
%		{\rm Im}\ T_{ab} = \frac{1}{2} \sum_{c} T_{ac} T_{bc}^* (2\pi)^D \delta^D(p_a-p_c) .
	\end{equation} 
%Defining $S$ by 
%$S = \textbf{1} + i T$ ($S^{\dagger} = \textbf{1} - i T^{\dagger}$), 
Then the unitarity relation reads that for any state $\Psi_a, \Psi_b \in \mathcal{V}$,
	\begin{align}
		{\rm Im} \langle \Psi_b |T| \Psi_a \rangle
		:=&   \frac{1}{2i} (\langle \Psi_b |T| \Psi_a \rangle - \langle \Psi_b |T^{\dagger}| \Psi_a \rangle) \nonumber\\
		=&  \frac{1}{2} \sum_{\Phi_n \in \mathcal{V}} \langle \Psi_b |T| \Phi_n \rangle \langle \Phi_n |T^{\dagger}| \Psi_a \rangle .
	\end{align}
\indent
On the other hand, the physical unitarity requires that  for any physical state $\Psi_a, \Psi_b \in \mathcal{V}_{\rm phys}$, only the physical states contribute to the intermediate states:

	\begin{equation}
		{\rm Im} \langle \Psi_b |T| \Psi_a \rangle
		= \frac{1}{2} \sum_{\Phi_n \in \mathcal{V}_{\rm phys}} \langle \Psi_b |T| \Phi_n \rangle \langle \Phi_n |T^{\dagger}| \Psi_a \rangle .
	\end{equation}
In other words, the physical unitarity  in gauge theories states that all the unphysical modes cancel in the intermediate states. 
The imaginary part is calculated by the Cutkosky cutting rule   \cite{Cutkosky60}.

\begin{figure}[tb]% 
%\vspace{30mm}
\begin{center}
\includegraphics[scale=0.22]{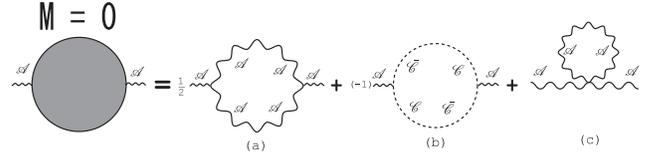}
\end{center}
%\capwidth90mm
\caption{
In the massless case $M=0$, the diagrams   contributing to the   amplitude $T(\mathscr{A} \rightarrow \mathscr{A})$ to the order $g^2$ are given by (a)~vector boson loop, (b)~ghost--antighost loop, (c)~boson tadpole.
}
\label{fig:unitarity-1}
\end{figure}

As a simple example, we consider a one-particle scattering, i.e., a  scattering process $\mathscr{A} \rightarrow \mathscr{A}$ in which the initial state is a massless gluon and the final state is also a massless gluon in the massless Yang-Mills theory as the $M=0$ case of the CF model. 
The Feynman rules in the $M=0$ case have been given in \cite{KMSI02}.
In the lowest order of the perturbation theory, the Feynman graphs of this process are given by Fig.~\ref{fig:unitarity-1}.  
Each diagram has one closed loop.
The initial state and the final state of a gluon are specified by the polarization vectors:
	\begin{equation}
		\varepsilon^{(\sigma) \alpha}(k) := \varepsilon^{\alpha} 
, \quad \varepsilon^{(\sigma) \alpha{}^\prime}(k) := \varepsilon^{\alpha{}^\prime}
 .
	\end{equation}

\begin{figure}[tb]%  
%\vspace{30mm}
\begin{center}
\includegraphics[scale=0.20]{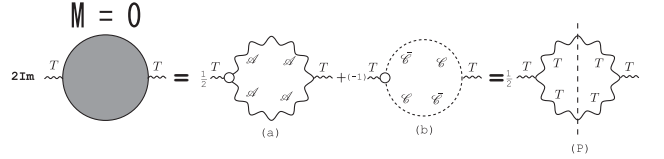}
\end{center}
%\capwidth90mm
\caption{
In the massless case $M=0$, the physical unitarity of the  amplitude $T(T^\pm \rightarrow T^\pm)$  to the order $g^2$  is checked according to the Cutkosky rule using the diagrams: (a)~vector boson loop, (b)~ghost--antighost loop.
}
\label{fig:unitarity-2}
\end{figure}

For this process up to the order $g^2$, we wish to check the physical unitarity relation for the scattering amplitude $T$.
By applying the Cutkosky rule  \cite{Cutkosky60} to Fig.~\ref{fig:unitarity-1}, we find that the imaginary part of the scattering amplitude $T(T^\pm \rightarrow T^\pm)$ from a transverse mode $T^\pm$ to a transverse mode $T^\pm$ is given by Fig.~\ref{fig:unitarity-2}. 
Here it should be remarked that the tadpole diagram does not have the imaginary part.

%%%%%%%%%%%%%%%%%%%%%%%%%%%%%%%%%%%%%%%%%%%%%%%%%%%%%%%%%%%%%%%%%%%
%%%%%%%%%%%%%%%%%%%%%%%%%%%%%%%%%%%%%%%%%%%%%%%%%%%%%%%%%%%%%%%%%%%

%%%%%%%%%%%%%%%%%%%%%%%%%%%%%%%%%%%%%%%%%%%%%%%%%%%%%%%%%%%%%%%%%%%
%%%%%%%%%%%%%%%%%%%%%%%%%%%%%%%%%%%%%%%%%%%%%%%%%%%%%%%%%%%%%%%%%%%	

The imaginary part of the diagram Fig.~\ref{fig:unitarity-2}(a) with a gluon loop is given by
	\begin{align}
		\frac{1}{2} &  (ig) f_{ABC} V_{\mu \nu \alpha}(k_1, k_2, k_3) \nonumber\\
		& \times 		(-ig) f_{ABC'} V_{\mu' \nu' \alpha'}(-k_1, -k_2, -k_3) \nonumber\\
		& \times   2\pi  \theta(k_1^0) \delta(k_1^2) g^{\mu \mu'} 2\pi \theta(k_2^0) \delta(k_2^2) g^{\nu \nu'}
\varepsilon^{\alpha}(k_3)\varepsilon^{\alpha'}(-k_3) ,
	\label{Fig2(a)0}
	\end{align}
where we have adopted the Feynman gauge $\beta=1$ for the gluon propagator and 
the factor 1/2 is the symmetrical factor due to two identical particles.
This  is written as 
	\begin{equation}
		  2g^2  \pi^2  f_{ABC} f_{ABC'} T_{\mu \nu} g^{\mu \mu'} g^{\nu \nu'} T_{\mu' \nu'} \theta(k_1^0)\delta(k_1^2) \theta(k_2^0)\delta(k_2^2) ,
	\label{Fig2(a)}
	\end{equation}
where we have defined
	\begin{align}
		&T_{\mu \nu} := V_{\mu \nu \alpha}(k_1, k_2, k_3) \varepsilon^{\alpha}(k_3) , \nonumber\\
		&T_{\mu' \nu'} := V_{\mu' \nu' \alpha'}(-k_1, -k_2, -k_3) \varepsilon^{\alpha'}(-k_3) .
	\end{align}

In the massless case $M = 0$, the physical unitarity requires the imaginary part of the sum of the second-order diagrams to be equal to 
	\begin{align}
	&	  2g^2  \pi^2 f_{ABC} f_{ABC'}  T_{\mu \nu} P^{\mu \mu'} P^{\nu \nu'} T_{\mu' \nu'}  \nonumber\\
		& \times \theta(k_1^0)\delta(k_1^2) \theta(k_2^0)\delta(k_2^2) 
		\nonumber\\
=&	\frac{1}{2}   (ig) f_{ABC} V_{\mu \nu \alpha}(k_1, k_2, k_3) \nonumber\\
& \times   		(-ig) f_{ABC'} V_{\mu' \nu' \alpha'}(-k_1, -k_2, -k_3) \nonumber\\
		& \times   2\pi  \theta(k_1^0) \delta(k_1^2) P^{\mu \mu'} 2\pi \theta(k_2^0) \delta(k_2^2) P^{\nu \nu'}
\varepsilon^{\alpha}(k_3)\varepsilon^{\alpha'}(-k_3) 
		\nonumber\\
=&	\frac{1}{2}   \sum_{a,b=1}^2 (ig  f_{ABC} V_{\mu \nu \alpha}(k_1, k_2, k_3)) \varepsilon_a^{\mu}(k_1) \varepsilon_b^{\nu}(k_2)  \varepsilon^{\alpha}(k_3)
	 \nonumber\\
		& \times
	( ig f_{ABC'} V_{\mu' \nu' \alpha'}(k_1, k_2, k_3))^* \varepsilon_a^{* \mu'}(k_1) \varepsilon_b^{* \nu'}(k_2) \varepsilon^{\alpha'}(-k_3)
 \nonumber\\
		& \times   2\pi  \theta(k_1^0) \delta(k_1^2)  2\pi \theta(k_2^0) \delta(k_2^2)  
 ,
	\label{Fig3}
	\end{align}
which is obtained from (\ref{Fig2(a)})  by the replacement:
	\begin{align}
		& g^{\mu \mu'} \to P^{\mu \mu'} = \sum_{a=1}^2 \varepsilon_a^{\mu}(k_1) \varepsilon_a^{* \mu'}(k_1) , \nonumber\\
		& g^{\nu \nu'} \to P^{\nu \nu'} = \sum_{b=1}^2 \varepsilon_b^{\nu}(k_2) \varepsilon_b^{* \nu'}(k_2) ,
	\label{Replace}
	\end{align}
where $P$ correspond to the two transverse polarization states for the massless spin-one modes $T^{\pm}$.

By using the decomposition:
	\begin{align}
		g^{\mu \mu'} &= - P^{\mu \mu'} + Q^{\mu \mu'}, \ Q^{\mu \mu'} = (k_1^{\mu} \bar{k}_1^{\mu'} + \bar{k}_1^{\mu} k_1^{\mu'})/(k_1 \cdot \bar{k}_1) , \nonumber\\
		g^{\nu \nu'} &= - P^{\nu \nu'} + Q^{\nu \nu'}, \ Q^{\nu \nu'} = (k_2^{\nu} \bar{k}_2^{\nu'} + \bar{k}_2^{\nu} k_2^{\nu'})/(k_2 \cdot \bar{k}_2) ,
	\end{align}
the difference between (\ref{Fig2(a)0})=(\ref{Fig2(a)}) and (\ref{Fig3}) is calculated from
	\begin{align}
		&g^{\mu \mu'} g^{\nu \nu'} - P^{\mu \mu'} P^{\nu \nu'} \nonumber\\
		=& - P^{\mu \mu'} Q^{\nu \nu'} - P^{\nu \nu'} Q^{\mu \mu'} + Q^{\mu \mu'} Q^{\nu \nu'} ,
	\label{Diff2}
	\end{align}
where $Q$ are rewritten using the polarization vectors for the longitudinal (L) and the scalar (S) modes:
%	\begin{equation}
%		P^{\mu \mu'} = \sum_{a=1}^2 \varepsilon_a^{\mu}(k_1) \varepsilon_a^{* \mu'}(k_1), \quad P^{\nu \nu'} = \sum_{b=1}^2 \varepsilon_b^{\nu}(k_2) \varepsilon_b^{* \nu'}(k_2) ,
%	\end{equation}
%and
	\begin{align}
		Q^{\mu \mu'} = \varepsilon_L^{\mu}(k_1) \varepsilon_S^{* \mu'}(k_1) + \varepsilon_S^{\mu}(k_1) \varepsilon_L^{* \mu'}(k_1) , \nonumber\\
		Q^{\nu \nu'} = \varepsilon_L^{\nu}(k_2) \varepsilon_S^{* \nu'}(k_2) + \varepsilon_S^{\nu}(k_2) \varepsilon_L^{* \nu'}(k_2).
	\end{align}
By using the relationships:
	\begin{align}
	&	T_{\mu \nu} \varepsilon_{\sigma}^{\mu}(k_1)\varepsilon_L^{\nu}(k_2) =
		T_{\mu \nu} \varepsilon_L^{\mu}(k_1)\varepsilon_{\sigma}^{\nu}(k_2) \nonumber\\\
&=
	\begin{cases}
	 (-i) k_{1\alpha} \varepsilon^{\alpha}(k_3)  \quad &(\sigma = S)    \\
	0   \quad &(\sigma = T,L)  
	\end{cases} ,
	\label{rel0}
	\end{align}
	and
	\begin{align}
	&	T_{\mu' \nu'} \varepsilon_{\sigma}^{*\mu'}(k_1)\varepsilon_L^{*\nu'}(k_2) =
		T_{\mu' \nu'} \varepsilon_L^{*\mu'}(k_1)\varepsilon_{\sigma}^{*\nu'}(k_2) \nonumber\\
&=
	\begin{cases}
	  i k_{1\alpha'} \varepsilon^{\alpha'}(-k_3)   \quad &(\sigma = S)    \\
	0   \quad &(\sigma = T,L)  
	\end{cases} ,
	\end{align}
we find that the first and second term in the right-hand side of (\ref{Diff2}) give vanishing contributions.

The relationship (\ref{rel0}) is derived as follows. 
First, we find
	\begin{align}
		&   V_{\mu \nu \alpha}(k_1, k_2, k_3) \varepsilon_L^{\nu}(k_2)  \nonumber\\
		=& V_{\mu \nu \alpha}(k_1, k_2, k_3) k^{\nu}_2  \nonumber\\
		=& (k_1-k_2)_{\alpha} k_{2\mu} + (k_2-k_3)_{\mu} k_{2\alpha} + (k_3\cdot k_2-k_1\cdot k_2) g_{\alpha\mu} \nonumber\\
%		=& -(2k_1+k_3)_{\alpha}(k_1+k_3)_{\mu} + (k_1+2k_3)_{\mu} (k_1+k_3)_{\alpha} + (k_1^2-k_3^2) g_{\alpha\mu} \nonumber\\
		=& - k_{1\alpha}k_{1\mu} + k_{3\alpha} k_{3\mu} 
%+ (k_1^2-k_3^2) g_{\alpha\mu}
 ,
	\label{V*L}
	\end{align}
where we have used $k_1^2=0$ and $k_3^2=0$ for the massless on-shell momenta.
Then we have
	\begin{align}
		&   V_{\mu \nu \alpha}(k_1, k_2, k_3)  \varepsilon_{\sigma}^{\mu}(k_1) \varepsilon_L^{\nu}(k_2) \varepsilon^{\alpha}(k_3) 
 \nonumber\\
		=& - k_{1\alpha} \varepsilon^{\alpha}(k_3) k_{1\mu} \varepsilon_{\sigma}^{\mu}(k_1) + k_{3\alpha} \varepsilon^{\alpha}(k_3) k_{3\mu} \varepsilon_{\sigma}^{\mu}(k_1) 
%+ (k_1^2-k_3^2) g_{\alpha\mu} \varepsilon_{\sigma}^{\mu}(k_1)
 ,
	\label{V*L*1}
	\end{align}
where 
$
k_{3\alpha} \varepsilon^{\alpha}(k_3)=0
$ 
for the physical transverse mode, 
while  
$
k_{1\mu} \varepsilon_{T}^{\mu}(k_1) = 0
$,
$
k_{1\mu} \varepsilon_{L}^{\mu}(k_1) = -ik_{1\mu} k_1^\mu= 0
$
for  massless on-shell $k_1$, 
and 
$
k_{1\mu} \varepsilon_{S}^{\mu}(k_1) = i
$
for  massless on-shell  $k_1$.

Thus, only the last term in (\ref{Diff2}) gives a nonvanishing contribution:
	\begin{align}
		& 4 g^2 \pi^2 f_{ABC} f_{ABC'} k_{1\alpha} \varepsilon^{\alpha}(k_3) k_{1\alpha'} \varepsilon^{\alpha'}(-k_3) \nonumber\\
	 & \times \theta(k_1^0)\delta(k_1^2) \theta(k_2^0)\delta(k_2^2) .
	\end{align}

This difference is exactly provided by the imaginary part of the second diagram of Fig.\ref{fig:unitarity-2}(b) with a  ghost loop:
	\begin{align}
  &    (-1)(-ig f_{ABC} k_{1\alpha}) (ig f_{ABC'}   k_{1\alpha'} )  \varepsilon^{\alpha}(k_3) \varepsilon^{\alpha'}(-k_3)  \nonumber\\
	 & \times 2\pi \theta(k_1^0)\delta(k_1^2) 2\pi \theta(k_2^0)\delta(k_2^2) 
\nonumber\\
	=&	- 4 g^2 \pi^2  f_{ABC} f_{ABC'} k_{1\alpha} \varepsilon^{\alpha}(k_3) k_{1\alpha'} \varepsilon^{\alpha'}(-k_3)  \nonumber\\
	 & \times  \theta(k_1^0)\delta(k_1^2) \theta(k_2^0)\delta(k_2^2) ,
	\end{align}
where we have used the original gluon-ghost-antighost vertex (given in \cite{KMSI02} at $\xi=1/2$ corresponding to  $\alpha=0$),
$
\frac12 ig f_{ABC} (k_{1\alpha}-k_{2\alpha})
= ig f_{ABC} (k_{1\alpha}+\frac12 k_{3\alpha})
$,
and the property of  polarization vectors,
$
k_{3\alpha} \varepsilon^{\alpha}(k_3)=0
$,
to obtain the first expression.	
Thus, the unitarity relation is satisfied in the massless case $M=0$.

%-------- Figure ----------------------------
\begin{figure}[t]
\begin{center}
\includegraphics[scale=0.25]{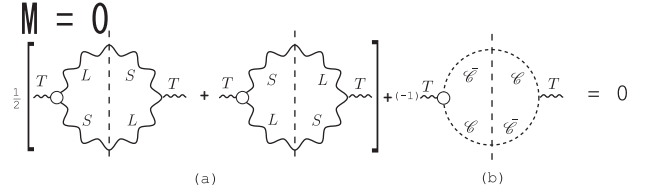}
\end{center}
\vspace{-0.3cm}
\caption{ 
In the massless case $M=0$, mode cancellations occur  to ensure the physical unitarity for the  one-particle  amplitude $T(T^\pm \rightarrow T^\pm)$ for the transverse mode $T$ to the order $g^2$. 
In the amplitude, two diagrams (a) from the longitudinal mode $L$ and the scalar mode $S$ are canceled by a ghost-antighost $\mathscr{C}, \mathscr{\bar{C}}$ diagram (b).  
%The NL model $\mathscr{N}$ is   non-propagating in the massless case and does not contribute to this cancellation. 
}
\label{fig:gauge-mode-cancel-massless}
\end{figure}
%----------------------------------------------

The physical unitarity  is ensured by  mode cancellations. 
See Fig.~\ref{fig:gauge-mode-cancel-massless}.
The contributions from the longitudinal mode $L$ and the scalar mode $S$ are canceled by a ghost-antighost $\mathscr{C}, \mathscr{\bar{C}}$ one.  
%It should be remarked that the NL model $N$ is non-propagating in the massless case and does not contribute to this cancellation. 

%%%%%%%%%%%%%%%%%%%%%%%%%%%%%%%%%%%%%%%%%%%%%%%%%%
%%%%%%%%%%%%%%%%%%%%%%%%%%%%%%%%%%%%%%%%%%%%%%%%%%
\section{Physical nonunitarity of massive Yang-Mills theory}
%%%%%%%%%%%%%%%%%%%%%%%%%%%%%%%%%%%%%%%%%%%%%%%%%%
%%%%%%%%%%%%%%%%%%%%%%%%%%%%%%%%%%%%%%%%%%%%%%%%%%

In this section, we reproduce the violation of physical unitarity in the massive Yang-Mills theory without the Higgs field based on the conventional argument, which shows the utility of the massive field $\mathscr{K}$ in discussing the physical unitarity of the CF model in the next section. 

In order to see a difference between massless gauge theory ($M=0$) and massive vector theory ($M \not= 0$), we consider the one-particle scattering, i.e., a scattering process $U \rightarrow U$ in which the initial state is a massive vector boson and the final state is also a massive vector boson. 
Here $U_\mu$ is defined by
	\begin{align}
		   U_\mu := \mathscr{A}_\mu - \frac{1}{M^2} \partial_\mu \mathscr{N} , \quad \text{or} \quad \mathscr{A}_\mu = U_\mu + \frac{1}{M^2} \partial_\mu \mathscr{N} .
	\end{align}

\begin{figure}[tb]%  
%\vspace{30mm}
\begin{center}
\includegraphics[scale=0.19]{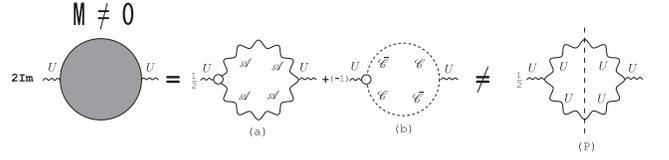}
\end{center}
%\capwidth90mm
\caption{
In the massive case $M \not= 0$, the physical unitarity of the  amplitude $T(U  \rightarrow U )$  for the physical spin-one vector mode $U$ to the order $g^2$ is checked according to the Cutkosky rule using the diagrams: (a) vector boson loop, (b) ghost--antighost loop.
The physical unitarity is violated due to the incomplete cancellation. 
}
\label{fig:unitarity-3}
\end{figure}

In the lowest order of the perturbation theory, the Feynman diagrams of this process are given by the same graphs as those in Fig.~\ref{fig:unitarity-1} where the propagators are replaced by the massive ones and the vertex functions are unchanged.  
Therefore, the imaginary part is given by the  diagrams of Fig.~\ref{fig:unitarity-3}.

The imaginary part of the diagram  Fig.~\ref{fig:unitarity-3}(a) with a loop of massive vector boson  is 
	\begin{align}
		& \frac{1}{2}   (-ig) f_{ABC} V_{\mu \nu \alpha}(k_1, k_2, k_3) \nonumber\\
		& \times   
		(ig) f_{ABC'} V_{\mu' \nu' \alpha'}(-k_1, -k_2, -k_3) \nonumber\\
		& \times   2\pi \theta(k_1^0)\delta(k_1^2-M^2) g^{\mu \mu'} \nonumber\\
		& \times   2\pi \theta(k_2^0)\delta(k_2^2-M^2) g^{\nu \nu'}  \varepsilon^{\alpha}(k_3)\varepsilon^{\alpha'}(-k_3) ,
	\label{Fig4(a)}
	\end{align}
where the factor 1/2 is the symmetrical factor due to two identical particles.
The imaginary part (\ref{Fig4(a)}) is written as 
	\begin{align}
		&  2 g^2 \pi^2  f_{ABC} f_{ABC'} T_{\mu \nu} g^{\mu \mu'} g^{\nu \nu'} T_{\mu' \nu'} \nonumber\\
		& \times   \theta(k_1^0)\delta(k_1^2 - M^2) \theta(k_2^0)\delta(k_2^2 - M^2) ,
	\label{Fig4(a)'}
	\end{align}
where we have defined
	\begin{align}
		&T_{\mu \nu} := V_{\mu \nu \alpha}(k_1, k_2, k_3) \varepsilon^{\alpha}(k_3) , \nonumber\\
		&T_{\mu' \nu'} := V_{\mu' \nu' \alpha'}(-k_1, -k_2, -k_3) \varepsilon^{\alpha'}(-k_3) .
	\end{align}

%%%%%%%%%%%%%%%%%%%%%%%%%%%%%%%%%%%%%%%%%%%%%%%%%%%%%%%%%%%%%%%%%%%
%%%%%%%%%%%%%%%%%%%%%%%%%%%%%%%%%%%%%%%%%%%%%%%%%%%%%%%%%%%%%%%%%%%

%-------- Figure ----------------------------
\begin{figure}[t]
\begin{center}
\includegraphics[scale=0.22]{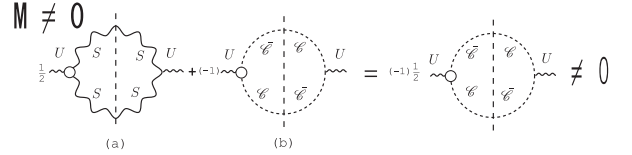}
\end{center}
\vspace{-0.3cm}
\caption{ 
In the massive case $M \not= 0$,   the incomplete mode cancellation among unphysical modes to the order $g^2$  prevents us from  ensuring the physical unitarity for the  one-particle  amplitude $T(U \rightarrow U)$ for the physical spin-one vector mode $U$. 
In the amplitude, a diagram  (a) from the scalar mode $S$ is overcanceled by a ghost-antighost  diagram (b), leaving one-half of (b) nonvanishing.  
}
\label{fig:gauge-mode-cancel}
\end{figure}
%----------------------------------------------
In the massive case $M \ne 0$, the physical unitarity requires the imaginary part of the second-order diagram to be equal to 
	\begin{align}
	&	  2 g^2 \pi^2 f_{ABC} f_{ABC'}  T_{\mu \nu} P^{\mu \mu'} P^{\nu \nu'} T_{\mu' \nu'} \nonumber\\
& \times \theta(k_1^0)\delta(k_1^2 - M^2) \theta(k_2^0)\delta(k_2^2 - M^2) 
\nonumber\\
 =&  \frac12  (igf_{ABC}  T_{\mu \nu} \varepsilon^{\mu}_j(k_1) \varepsilon^{\nu}_\ell(k_2))
\nonumber\\
&  \times (igf_{ABC'} T_{\mu' \nu'} \varepsilon^{\mu'}_j(k_1) \varepsilon^{\nu'}_\ell(k_2))^*
\nonumber\\
& \times 2\pi \theta(k_1^0)\delta(k_1^2 - M^2) 2\pi \theta(k_2^0)\delta(k_2^2 - M^2) 
 ,
	\label{Fig3'}
	\end{align}
which is obtained from (\ref{Fig4(a)'}) by the replacement:
	\begin{align}
	&	g^{\mu \mu'} \to P^{\mu \mu'} = \sum_{j=1}^3 \varepsilon^{\mu}_{(j)}(k_1) \varepsilon^{\mu'}_{(j)}(k_1), \nonumber\\
	&	g^{\nu \nu'} \to P^{\nu \nu'} = \sum_{j=1}^3 \varepsilon^{\nu}_{(j)}(k_2) \varepsilon^{\nu'}_{(j)}(k_2).
	\label{Replace2}
	\end{align}
	
Using the decomposition for $M \not= 0$:
	\begin{equation}
		g^{\mu \mu'} = - P^{\mu \mu'} + \frac{k_1^{\mu} k_1^{\mu'}}{M^2} , \quad
		g^{\nu \nu'} = - P^{\nu \nu'} + \frac{k_2^{\nu} k_2^{\nu'}}{M^2} ,
	\end{equation}
where $P$ selects three polarization states ($T^+, T^-, L$) for the massive spin-one boson $U$,  
the difference between (\ref{Fig4(a)})=(\ref{Fig4(a)'}) and (\ref{Fig3'}) is calculated from
	\begin{align}
		&  g^{\mu \mu'} g^{\nu \nu'} - P^{\mu \mu'} P^{\nu \nu'} 	\nonumber\\
 =&   - P^{\mu \mu'} \frac{k_2^{\nu} k_2^{\nu'}}{M^2} - P^{\nu \nu'} \frac{k_1^{\mu} k_1^{\mu'}}{M^2} + \frac{k_1^{\mu} k_1^{\mu'}}{M^2} \frac{k_2^{\nu} k_2^{\nu'}}{M^2} .
	\label{Diff}
	\end{align}
The contribution from the first term of the right-hand side of (\ref{Diff}) to (\ref{Fig4(a)'})  is given by 
	\begin{align}
	&	2 g^2 \frac{1}{M^2}  \pi^2 f_{ABC} f_{ABC'} T_{\mu \nu} \varepsilon^{\mu}_j(k_1) k_2^{\nu} \varepsilon^{\mu'}_j(k_1) k_2^{\nu'} T_{\mu' \nu'}  \nonumber\\&
\times \theta(k_1^0)\delta(k_1^2 - M^2) \theta(k_2^0)\delta(k_2^2 - M^2) .
	\label{GhostCont}
	\end{align}
This is zero, since
	\begin{align}
	&	T_{\mu \nu}  \varepsilon^{\mu}_j(k_1) k_2^{\nu} \nonumber\\
		&= \varepsilon^{\mu}_j(k_1) V_{\mu \nu \alpha}(k_1, k_2, k_3) k_2^{\nu}  \varepsilon^{\alpha}(k_3) \nonumber\\
		&= \varepsilon^{\mu}_j(k_1) [- k_{1 \mu} k_{1 \alpha} + k_{3\mu}k_{3\alpha} + k_1^2g_{\mu \alpha} - k_3^2 g_{\mu \alpha}] \varepsilon^{\alpha}(k_3) \nonumber\\
		&=  - k_{1 \mu} \varepsilon^{\mu}_j(k_1) k_{1 \alpha} \varepsilon^{\alpha}(k_3) + k_{3\mu} \varepsilon^{\mu}_j(k_1) k_{3\alpha} \varepsilon^{\alpha}(k_3)   \nonumber\\
		&= 0 ,
	\end{align}
where we have used $k_1^2=k_3^2$ and $k_{\mu} \varepsilon^{\mu}_{(j)}(k)=0$.
Similarly, the contribution from the second term of (\ref{Diff}) is vanishing.
The third term of (\ref{Diff}) gives
	\begin{equation}
		 2 g^2  \pi^2 f_{ABC} f_{ABC'} T_{\mu \nu} \frac{k_1^{\mu} k_2^{\nu}}{M^2} \frac{k_1^{\mu'} k_2^{\nu'}}{M^2} T_{\mu' \nu'} .
	\end{equation}
Using the property:
	\begin{align}
		T_{\mu \nu} k_1^{\mu} k_2^{\nu}
		&=  - k_1^2 k_{1 \alpha} \varepsilon^{\alpha}(k_3) + k_1 \cdot k_3 k_{3\alpha} \varepsilon^{\alpha}(k_3)  \nonumber\\
		&=  - M^2 k_{1 \alpha} \varepsilon^{\alpha}(k_3) ,
	\end{align}
 we obtain the difference:
	\begin{align}
		&   2 g^2  \pi^2 f_{ABC} f_{ABC'} k_{1 \alpha} \varepsilon^{\alpha}(k_3) k_{1 \alpha'} \varepsilon^{\alpha'}(-k_3) \nonumber\\
		& \times \theta(k_1^0)\delta(k_1^2 - M^2) \theta(k_2^0)\delta(k_2^2 - M^2) .
		\label{VectorCont2}
	\end{align}

This difference must be provided by the imaginary part of the diagram Fig.~\ref{fig:unitarity-3}(b) with a loop of massive ghost  which is given by
	\begin{align}
		&(-1) \varepsilon^{\alpha}(k_3) \varepsilon^{\alpha'}(-k_3) (-ig f_{ABC} k_{1 \alpha}) (ig f_{ABC'} (-k_{2 \alpha'})) \nonumber\\& 
\times 2\pi \theta(k_1^0)\delta(k_1^2 - M^2)  2\pi \theta(k_2^0)\delta(k_2^2 - M^2)
 \nonumber\\
		=& -4 g^2 \pi^2 f_{ABC} f_{ABC'} k_{1 \alpha} \varepsilon^{\alpha}(k_3) k_{1 \alpha'} \varepsilon^{\alpha'}(-k_3) \nonumber\\&
\times \theta(k_1^0)\delta(k_1^2 - M^2) \theta(k_2^0)\delta(k_2^2 - M^2) ,
		\label{GhostCont2}
	\end{align}
where we have used $k_2=-k_1-k_3$ and $k_{3 \alpha} \varepsilon^{\alpha}(k_3) = 0$.
The ghost contribution (\ref{GhostCont2}) is precisely of the same form as (\ref{VectorCont2}) and comes with the opposite sign.
However, the massive vector contribution (\ref{VectorCont2}) cancels against half the massive ghost contribution (\ref{GhostCont2}).
See Fig.\ref{fig:gauge-mode-cancel}.

Thus, it is found that there is a discrete difference between massless theories and massive theories.
This means that the massless theory cannot be obtained as a limiting case of the massive theory.
The origin of the difference goes back to the difference between the sum over polarizations. %\cite{vDV70} 
In the next section, we re-examine the physical unitarity in the the CF model based on our method.

%%%%%%%%%%%%%%%%%%%%%%%%%%%%%%%%%%%%%%%%%%%%%%%%%%%%%%%%%%%%%%%%%%%%%%%%
%%%%%%%%%%%%%%%%%%%%%%%%%%%%%%%%%%%%%%%%%%%%%%%%%%%%%%%%%%%%%%%%%%%%%%%%
\section{Perturbative analysis of physical unitarity in the CF model}
%%%%%%%%%%%%%%%%%%%%%%%%%%%%%%%%%%%%%%%%%%%%%%%%%%%%%%%%%%%%%%%%%%%%%%%%
%%%%%%%%%%%%%%%%%%%%%%%%%%%%%%%%%%%%%%%%%%%%%%%%%%%%%%%%%%%%%%%%%%%%%%%%

In order to re-examine the physical unitarity in the massive Yang-Mills theory, i.e., the CF model,
we consider the simplest case of the one-particle amplitude $T(\mathscr{K} \rightarrow \mathscr{K})$  in the perturbation theory, as considered in \cite{DV70}.

\begin{figure}[t]
\begin{center}
\includegraphics[scale=0.20]{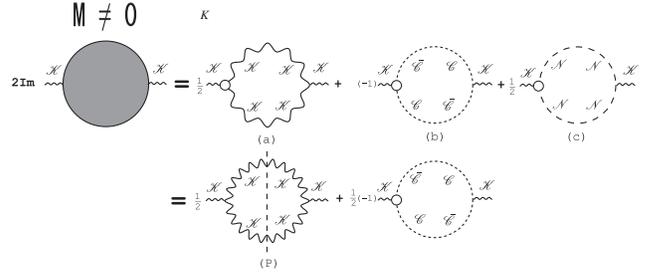}
\end{center}
\vspace{-0.3cm}
\caption{ 
In the massive case $M \not= 0$, the physical unitarity of the  amplitude $T(\mathscr{K} \rightarrow \mathscr{K})$ to the order $g^2$  is checked according to the Cutkosky rule using the diagrams: (a)~massive vector boson, (b)~ghost--antighost, (c)~NL field.
}
\label{fig:diagram}
\end{figure}

According to the Cutkosky rules, the physical unitarity up to the order $g^2$ is checked by calculating the diagrams in Fig.~\ref{fig:diagram} with one closed loop.

The imaginary part for the diagram   Fig.~\ref{fig:diagram}(a) is 
	\begin{align}
		(\rm{a})_{\alpha \alpha'}^{CC'} :=& \left( \frac{1}{2} \right) (ig) f_{ABC} V^{\mu\nu\alpha}(k_1, k_2, k_3) 
\nonumber\\
		&\times (-ig) f_{A'B'C'} V^{\mu'\nu'\alpha'}(-k_1, -k_2, -k_3) \nonumber\\
		&\times  2\pi \theta(k_1^0)  \delta^{AA'} I_{\mu\mu'}(k_1)  2\pi \theta(k_2^0)  \delta^{BB'} I_{\nu\nu'}(k_2) ,
	\end{align}
where we have defined
%	\begin{equation}
%		V_{\mu\nu\alpha}(k_1, k_2, k_3) = (k_1-k_2)_{\alpha}g_{\mu\nu} + (k_2-k_3)_{\mu}g_{\nu\alpha} + (k_3-k_1)_{\nu}g_{\alpha\mu} ,
%	\end{equation}
%and
	\begin{align}
     I_{\mu\nu}(k) 
:=&    \delta(k^2-M^2) \left( g_{\mu\nu} -\frac{k_\mu k_\nu}{M^2} \right) 
%\nonumber\\
%&+ 2\pi \theta(k^0) 
%+ \delta(k^2-\beta M^2)  \frac{k_\mu k_\nu}{M^2}  
 .
    \end{align}
    Remember that the three polarization vectors $\varepsilon_{\mu}^{(j)}(k) (j=1,2,3)$ for the spin-one massive vector field $\mathscr{K}_{\mu}$ obey the relation:
	\begin{align}
		\sum_{j=1}^3 \varepsilon_{\mu}^{(j)}(k_1) \varepsilon_{\mu'}^{(j)*}(k_1) = - g_{\mu \mu'} + \frac{k_{1\mu}k_{1\mu'}}{M^2}, \nonumber\\
		\sum_{j=1}^3 \varepsilon_{\nu}^{(j)}(k_2) \varepsilon_{\nu'}^{(j)*}(k_2) = - g_{\nu \nu'} + \frac{k_{2\nu}k_{2\nu'}}{M^2} .
	\end{align}
Therefore,  (a) is a physical contribution coming from the spin-one massive vector boson $\mathscr{K}$.
The physical unitarity requires that the contributions other than (a), i.e.,  the contributions (b) and (c) from the unphysical fields $\mathscr{C}, \bar{\mathscr{C}}$ and $\mathscr{N}$ are canceled  in the same order of the coupling.
Therefore, we consider the contributions from unphysical fields.

%-------- Figure ----------------------------
%----------------------------------------------

%-------- Figure ----------------------------
\begin{figure}[t]
\begin{center}
\includegraphics[scale=0.22]{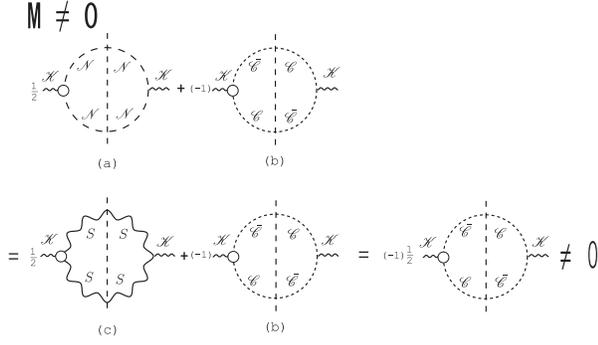}
\end{center}
\vspace{-0.3cm}
\caption{ 
In the massive case  $M \not= 0$, incomplete mode cancellations violate the physical unitarity for the  one-particle  amplitude to the order $g^2$. 
%The top diagram corresponds to the massless case $M=0$. 
%The amplitude $T \rightarrow T$ for the transverse mode $T$, two diagrams (a)  from the longitudinal model $L$ and the scalar mode $S$ are canceled by a ghost-antighost $C, \bar{C}$ diagram (b). 
%The NL model $\mathscr{N}$ is   non-propagating in the massless case and does not contribute to this cancellation. 
%$$
% \boxed{ T1, T2}, L, S; C, C; (N: {\rm non-propagating})
%$$
%The bottom diagram corresponds to the massive case $M \not= 0$. 
%The amplitude $K \rightarrow K$ for a physical particle $K$, a diagram (a) from the scalar mode is not sufficient to cancel the ghost-antighost $C, \bar{C}$ diagram (b). An additional diagram (c) from the NL model $\mathscr{N}$ is necessary to cancel. 
%$$
% \boxed{ T1, T2, L}, S;  C ,   C; N 
%$$
%The NL model $\mathscr{N}$ is propagating and plays the important role in the cancellation in the massive case, in sharp contrast to the massless case.  
}
\label{fig:gauge-mode-cancel-2}
\end{figure}
%----------------------------------------------

The imaginary part for the diagram Fig.~\ref{fig:diagram}(b) with a closed loop of the massive FP ghost and antighost reads
	\begin{align}
	&	(b)_{\alpha \alpha'}^{CC'} 
\nonumber\\	 :=& (-1) (-ig) f_{ABC} \{ M^{-2} V_{\alpha}(k_1, k_2, k_3) + k_{1\alpha} - k_{2\alpha} \} 
\nonumber\\
		&\times ig f_{A'B'C'} \{ M^{-2} V_{\alpha'}(k_1, k_2, k_3) + k_{1\alpha'} - k_{2\alpha'} \} 
\nonumber\\
		&\times		  2\pi \theta(k_1^0) \delta(k_1^2-\beta M^2) 2\pi \theta(k_2^0) \delta(k_2^2-\beta M^2) \nonumber\\
		=& -4 g^2 \pi^2 f_{ABC} f_{ABC'} \{ M^{-2} V_{\alpha}(k_1, k_2, k_3) + k_{1\alpha} - k_{2\alpha}  \} \nonumber\\
		& \times	\{ \alpha \to \alpha'\} 
  \theta(k_1^0) \delta(k_1^2-\beta M^2) \theta(k_2^0) \delta(k_2^2-\beta M^2) .
	\end{align}

The imaginary part of the diagram (c) with a closed loop of the NL field reads
	\begin{align}
	&	(c)_{\alpha \alpha'}^{CC'}
\nonumber\\		
:= & \left(\frac{1}{2} \right) (-i) \frac{g}{M^2} f_{ABC} \{ M^{-2} V_{\alpha}(k_1, k_2, k_3) + k_{1\alpha} - k_{2\alpha} \} \nonumber\\
		&\times i \frac{g}{M^2} f_{A'B'C'} \{ M^{-2} V_{\alpha'}(k_1, k_2, k_3) + k_{1\alpha'} - k_{2\alpha'}  \} \nonumber\\
		&\times M^2 2\pi  \theta(k_1^0) \delta(k_1^2-\beta M^2) \delta^{AA'} M^2 \nonumber\\
		&\times	2\pi  \theta(k_2^0) \delta(k_2^2-\beta M^2) \delta^{BB'} \nonumber\\
		=&   2 g^2 \pi^2 f_{ABC} f_{ABC'} \{ M^{-2} V_{\alpha}(k_1, k_2, k_3) + k_{1\alpha} - k_{2\alpha} \} \nonumber\\
		&\times	\{  \alpha \to \alpha' \} 
  \theta(k_1^0) \delta(k_1^2-\beta M^2) \theta(k_2^0) \delta(k_2^2-\beta M^2) .
	\end{align}
Hence, adding the NL loop (c) to the ghost loop (b) yields the half of (b): $(b) + (c) = \frac{1}{2} (b)$, since $(c) = -\frac{1}{2} (b)$.
Therefore,  we have shown the CF model does not satisfy the physical unitarity for the $M \not= 0$ case independently from  $\beta$ in the perturbation theory.
The incomplete cancellations of the unphysical modes against the physical unitarity are summarized in Fig.~\ref{fig:gauge-mode-cancel-2}.
This result should be compared with the massless case where  the physical modes  in the massless case $M=0$ are the two transverse parts.

In the massless case $M=0$, 
for the amplitude $T \rightarrow T$ of the transverse mode  $T^{\pm}$, two diagrams (a)  from the longitudinal mode  $L$ and the scalar mode $S$ are canceled by a ghost-antighost $\mathscr{C} , \mathscr{\bar C}$ diagram (b). 
The NL mode  $\mathscr{N}$ is   nonpropagating in the massless case and does not contribute to this cancellation.

In the massive case $M \not= 0$, 
for the amplitude $\mathscr{K} \rightarrow \mathscr{K}$ of a physical particle $\mathscr{K}$, a diagram (a) from the scalar mode $S$ (= the NL mode $\mathscr{N}$) is not sufficient to cancel the ghost-antighost $\mathscr{C} , \mathscr{\bar C}$ diagram (b). An additional bosonic contribution is necessary to realize the complete cancellation. 
%The NL model $\mathscr{N}$ is propagating and could play the important role in the cancellation in the massive case, in sharp contrast to the massless case.  ???

%%%%%%%%%%%%%%%%%%%%%%%%%%%%%%%%%%%%%%%%%%%%%%%%%%%%%%%%%%%%%%%%%%%%%%%%
%%%%%%%%%%%%%%%%%%%%%%%%%%%%%%%%%%%%%%%%%%%%%%%%%%%%%%%%%%%%%%%%%%%%%%%%
%\section{How to avoid the violation of  physical unitarity}
%%%%%%%%%%%%%%%%%%%%%%%%%%%%%%%%%%%%%%%%%%%%%%%%%%%%%%%%%%%%%%%%%%%%%%%%
%%%%%%%%%%%%%%%%%%%%%%%%%%%%%%%%%%%%%%%%%%%%%%%%%%%%%%%%%%%%%%%%%%%%%%%%

In the massless case $M=0$, the physical modes are given by two  transverse modes $T^{\pm}$.
Then the two unphysical modes in the gluon, i.e.,  the longitudinal mode  $L$ and the scalar mode $S$ are canceled by a ghost and antighost $\mathscr{C} , \mathscr{\bar C}$. 
Two bosonic modes are exactly canceled by two fermionic (anticommuting) modes. 
It should be remarked that the scalar mode is identified with the NL mode  $\mathscr{N}$, which is non-propagating in the massless case. 
\begin{equation}
 \fbox{ $T^+$, $T^-$}, L, S (=\mathscr{N}: {\rm nonpropagating}); \mathscr{C} , \mathscr{\bar C}  
\end{equation}

In the massive case $M \not= 0$, the physical modes are given by a longitudinal and two transverse modes. 
A remaining unphysical mode, i.e., a scalar mode is not sufficient to cancel the ghost and antighost contributions. 
Therefore, the elementary fields in the original action of the CF model are not sufficient to respect the physical unitarity. 
%An additional (bosonic) mode must be supplied by some reasons. 
There must be a mechanism which supplies the CF model with an extra (bosonic) mode.
(Note that the NL field  $\mathscr{N}$ is propagating in the massive case and therefore  the NL mode is expected to play the important role in the cancellation in the massive case, in sharp contrast to the massless case.  However, the NL mode  is identical to the scalar mode on mass shell and hence cannot be counted as another independent field.) 
\begin{equation}
 \fbox{ $T^+$, $T^-$, $L$}, S=\mathscr{N};  \mathscr{C} , \mathscr{\bar C} . 
\end{equation}

Finally, we discuss how to avoid the violation of physical unitarity.
The violation of the physical unitarity is avoided by restricting the relevant energy   to the low-energy region such that the ghost and antighost pair cannot be created.  
This can be done by adjusting the parameter in the CF model. 
Since the ghost and antighost have the same mass $\sqrt{\beta}M$, the allowed region is 
\begin{equation}
    E < 2 \sqrt{\beta}M , \quad E^2 < 4\beta M^2.
\end{equation}	  	
A shortcoming of this scenario is that $\beta=0$ is not allowed to maintain physical unitarity, since the results of numerical simulations on the lattice are available only in this case $\beta=0$. 

At first glance, the cancellation of unphysical modes works well even in the massive case by using the argument similar to that done in the gauge-Higgs model with the renormalizable $R_\xi$ gauge. 
The pole masses of unphysical fields $\mathscr{C}, \mathscr{\bar C}, \mathscr{N}$ are the same:
\begin{align}
 m_{\mathscr{C}}^2 = m_{\mathscr{\bar C}}^2 = m_{\mathscr{N}}^2 = \beta M^2.
\end{align}
In the limit $\beta \rightarrow \infty$, 
unphysical fields $\mathscr{C}, \mathscr{\bar C}, \mathscr{N}$ decouple from the theory, leaving the physical field $\mathscr{K}_\mu$ in the theory.
Therefore, it seems that the physical unitarity holds even in the massive case. 
However, this is not the case, as we have shown in the above. 
What is wrong in this argument?
This argument is based on the fact that $\beta$ is a gauge-fixing parameter and the physics does not depend on this parameter $\beta$, which is indeed shown in the massless case $M=0$. 
However, even the BRST-invariant quantities depend on $\beta$ for $M \not =0$. 
Therefore, the physics depends on $\beta$ for $M \not =0$, and  the physics for $\beta \rightarrow \infty$ is different from that for $\beta < \infty$.
In this sense, the above result, i.e., violation of physical unitarity in the massive case, does not contradict   this argument.

In order to maintain the physical unitarity in the massive Yang-Mills theory without the Higgs field, therefore, we need a  nonperturbative approach, which will be given in subsequent papers in preparation.

\section{Conclusion and discussion}

In order to understand color confinement in QCD in the light of recent developments, we have considered a ``massive'' Yang-Mills model without the Higgs field, especially, the CF model, 
since the CF model is regarded as a good low-energy effective theory of QCD  and it is much simpler than the refined Gribov-Zwanziger model, see e.g., \cite{Kondo12}.

%In the quantum field theory for the massive vector boson, it is believed that physical unitarity and  renormalizablity are compatible with each other only when the massive vector boson is provided by the Higgs mechanism where the gauge symmetry is spontaneously broken by the existence of the Higgs field.

We have examined the physical unitarity of  the  CF model which is known to be renormalizable.
%which is expected to be blessed with physical unitarity and renormalizability .
For this purpose, we have used the field $\mathscr{K}_\mu^A(x)$ with the following properties: 
\begin{enumerate}
\item[
(i)]
 $\mathscr{K}$ is invariant  under an extended BRST transformation, ${\boldsymbol \delta}' \mathscr{K}_\mu(x)=0$ (off shell).

\item[
(ii)] $\mathscr{K}$ is divergenceless,   
$\partial^\mu \mathscr{K}_\mu(x)=0$ (on shell).

\item[
(iii)] $\mathscr{K}$ transforms according to the adjoint representation under color rotation.
$\mathscr{K}_{\mu}(x) \to U \mathscr{K}_{\mu}(x) U^{-1}$

\item[
(iv)]
 $\mathscr{K}$ is invariant under the FP conjugation.
$\mathscr{K}_{\mu}(x) \to \mathscr{K}_{\mu}(x)$ 
\end{enumerate}
\noindent
Thus, we have identified $\mathscr{K}_\mu^A$ with a physical and massive vector field with correct degrees of freedom  as a non-Abelian spin-one massive vector boson.
$\mathscr{K}$ is obtained by a nonlinear but local transformation from the original fields in the CF model.

We  have checked in a new perturbative treatment whether or not the CF model satisfies the physical unitarity.
Then we have confirmed the violation of the physical unitarity in the perturbative treatment and we have clarified the reason  in the massive Yang-Mills theory without the Higgs field. 
The perturbative analysis for the   physical unitarity imposea a restriction on the valid energy together with  the parameter of the CF model:
$E^2 < 4\beta M^2$ 
in order to confine unphysical modes (ghost, antighost, scalar mode).
However, $\beta=0$ is not allowed in this scenario.

The conclusion  obtained in this paper still leaves  a possibility that the nonperturbative approach can modify the conclusion.
In a subsequent paper, indeed, we will propose  a scenario in which the physical unitarity can be recovered in the CF model thanks to the FP conjugation invariance. 
Indeed, we will show that the norm cancellation is automatically guaranteed from the Slavnov-Taylor identities if the ghost-antighost bound state exists. 
In this way, the physical unitarity can be recovered in a nonperturbative way. 
To show the existence of the bound state of ghost and antighost,  the Nambu-Bethe-Salpeter equation is to be solved. This is a hard work to be tackled in subsequent papers.

{\it Acknowledgements}:\ 
The authors would like to thank Professor T. Kugo for very enlightening discussion about some issues. 
This work is  supported by Grant-in-Aid for Scientific Research (C)  24540252 from the Japan Society for the Promotion of Science (JSPS).

\appendix
%%%%%%%%%%%%%%%%%%%%%%%%%%%%%%%%%%%%%%%%%%%%%%%%%%%%%%%%
%%%%%%%%%%%%%%%%%%%%%%%%%%%%%%%%%%%%%%%%%%%%%%%%%%%%%%%%
\section{Change of variables}
%%%%%%%%%%%%%%%%%%%%%%%%%%%%%%%%%%%%%%%%%%%%%%%%%%%%%%%%
%%%%%%%%%%%%%%%%%%%%%%%%%%%%%%%%%%%%%%%%%%%%%%%%%%%%%%%%

The original theory is given by
	\begin{equation}
	Z_{\rm mYM} = \int \mathcal{D}\mathscr{A}  \mathcal{D}\mathscr{C}  \mathcal{D}\mathscr{\bar C}  \mathcal{D}\mathscr{N} e^ { iS^{\rm tot}_{\rm mYM}[\mathscr{A},\mathscr{C},\mathscr{\bar C},\mathscr{N}] } .
	\end{equation}
The exact change of variables, $\mathscr{A}_\mu, \mathscr{C}, \bar{\mathscr{C}}, \mathscr{N} \rightarrow \mathscr{K}_\mu, \mathscr{C}, \bar{\mathscr{C}}, \mathscr{N}$, 
could be performed through the relationship $\mathscr{A}= \mathscr{\tilde A}[\mathscr{K}_\mu, \mathscr{C}, \bar{\mathscr{C}}, \mathscr{N}]$ according to 
	\begin{align}
	Z_{\rm mYM} 
&= \int \mathcal{D}\mathscr{K}  \mathcal{D}\mathscr{C}  \mathcal{D}\mathscr{\bar C}  \mathcal{D}\mathscr{N} \Big|\frac{\partial(\mathscr{\tilde A})}{\partial(\mathscr{K})}\Big| 
e^{ iS^{\rm tot}_{\rm mYM}[\mathscr{\tilde A},\mathscr{C},\mathscr{\bar C},\mathscr{N}] } 
\nonumber\\
&= \int \mathcal{D}\mathscr{K}  \mathcal{D}\mathscr{C}  \mathcal{D}\mathscr{\bar C}  \mathcal{D}\mathscr{N}   e^{ iS^{\rm tot}_{\rm mYM}[\mathscr{K},\mathscr{C},\mathscr{\bar C},\mathscr{N}]  } 
,
	\end{align}
where 
	\begin{equation}
	S^{\rm tot}_{\rm mYM}[\mathscr{K},\mathscr{C},\mathscr{\bar C},\mathscr{N}] :=  S^{\rm tot}_{\rm mYM}[\mathscr{\tilde A},\mathscr{C},\mathscr{\bar C},\mathscr{N}] -i \ln J ,
	\end{equation}
	where $J$ is the Jacobian  associated with the change of variables from $\mathscr{A}_\mu, \mathscr{C}, \bar{\mathscr{C}}, \mathscr{N}$ in the original Lagrangian to $\mathscr{K}_\mu, \mathscr{C}, \bar{\mathscr{C}}, \mathscr{N}$ in the new theory.

We proceed to check the modified BRST (mBRST) invariance of the new theory. 
The action is mBRST invariant by construction. 
We have already shown that the integration measure $\mathcal{D}\mathscr{A}  \mathcal{D}\mathscr{C}  \mathcal{D}\mathscr{\bar C}  \mathcal{D}\mathscr{N}$ is mBRST invariant.  The mBRST invariance of the measure $\mathcal{D}\mathscr{K}  \mathcal{D}\mathscr{C}  \mathcal{D}\mathscr{\bar C}  \mathcal{D}\mathscr{N}$ is also checked in the same way.  
Therefore, the Jacobian $J$ must be mBRST invariant too, i.e., ${\boldsymbol \delta}'J=0$. 

However, we do not know the exact expression of $\mathscr{A}= \mathscr{\tilde A}[\mathscr{K}_\mu, \mathscr{C}, \bar{\mathscr{C}}, \mathscr{N}]$ as a functional of $\mathscr{K}_\mu, \mathscr{C}, \bar{\mathscr{C}}, \mathscr{N}$, although we know the exact expression of $\mathscr{K}= \mathscr{\tilde K}[\mathscr{A}_\mu, \mathscr{C}, \bar{\mathscr{C}}, \mathscr{N}]$ as a functional of $\mathscr{A}_\mu, \mathscr{C}, \bar{\mathscr{C}}, \mathscr{N}$ as given in (\ref{K}). 
We know just the order by order relation for $\mathscr{A}= \mathscr{\tilde A}[\mathscr{K}_\mu, \mathscr{C}, \bar{\mathscr{C}}, \mathscr{N}]$ as given in (\ref{A_mu-expan}).
Hence we can calculate the Jacobian $J$ order by order of the coupling constant $g$. 	
Thus, the mBRST invariance of $J$  can be checked order by order in the coupling constant $g$, although  the full mBRST invariance cannot be checked because we do not know the exact form of $J$. 
Here it should be remarked that ${\boldsymbol \delta}' \mathscr{N}=M^2 \mathscr{C}$ and ${\boldsymbol \delta}' \mathscr{\bar C}=i\mathscr{N}$ do not change the order, while ${\boldsymbol \delta}' \mathscr{C}=-\frac12 g (\mathscr{C} \times \mathscr{C})$ and ${\boldsymbol \delta}' \mathscr{A}_\mu=\mathscr{D}_\mu[\mathscr{A}]\mathscr{C}$ increase the order of $g$ by one.

In fact, the Jacobian $J$  is calculated as follows. 
The integration measure is transformed as
	\begin{align}
	& \mathcal{D} \mathscr{A} \mathcal{D} \mathscr{C} \mathcal{D} \bar{\mathscr{C}} \mathcal{D} \mathscr{N}  =	   \mathcal{D} \mathscr{K} \mathcal{D} \mathscr{C}' \mathcal{D} \bar{\mathscr{C}}' \mathcal{D} \mathscr{N}' 	\ J	  .
	\end{align}
The Jacobian $J$ is calculated as
%[Exercise-10] \marginpar{Ex-10}
%\small
	\begin{align}
  J^{-1} =& {\rm Det}
		\left[ \frac{\delta \mathscr{K}_{\mu}^{A}(x)}{\delta \mathscr{A}_{\nu}^B(y)} \right] 
 \nonumber\\		 
		=&  {\rm Det}
		  \Big[ \delta_\mu^\nu (\delta^{AB} - g M^{-2}f^{ABC}\mathscr{N}^C 
\nonumber\\
 & \quad\quad + g^2M^{-2}f^{ACE}f^{BFE} i\bar{\mathscr{C}}^C \mathscr{C}^F) \delta^D(x-y) \Big]
  ,
	\end{align}
where we have used
	\begin{align}
 \mathscr{K}_{\mu}^A  =& \mathscr{A}_{\mu}^A
		- \frac{1}{M^2} \partial_{\mu} \mathscr{N}^A
		\nonumber\\
		&  - \frac{g}{M^2} f^{ABC} \mathscr{A}_{\mu}^B   \mathscr{N}^C + \frac{g}{M^2} f^{ABC} \partial_{\mu}\mathscr{C}^B  i\bar{\mathscr{C}}^C
\nonumber\\
		&+ \frac{g^2}{M^2} f^{AEC}f^{EGF}  \mathscr{A}_{\mu}^G  \mathscr{C}^F  i \bar{\mathscr{C}}^C .
	\label{K_mu2}
	\end{align}
Applying the formula: 
\begin{align}
& {\rm Det}(1+X) 
\nonumber\\
=&  \exp [{\rm tr} \ln (1+X) ]
\nonumber\\
=& \exp \left[ {\rm tr}(X)-\frac12 {\rm tr}(X^2)+\frac13 {\rm tr}(X^3) - \frac14 {\rm tr}(X^4) + ... \right] ,
\end{align}
to
\begin{align}
 X^{AB} = - g M^{-2}f^{ABC}\mathscr{N}^C  + g^2M^{-2}f^{ACE}f^{BFE} i\bar{\mathscr{C}}^C \mathscr{C}^F .
\end{align}
we find that the order $g$ contribution (from the NL field) vanishes  due to  $f^{AAC}=0$:
	\begin{align}
  J^{-1} =&  \prod_{x} \exp \Big[   \ 
    \frac12 g^2 M^{-4} N_c \mathscr{N} \cdot \mathscr{N}
+  g^2M^{-2}N_c i\bar{\mathscr{C}} \cdot \mathscr{C}  
\nonumber\\
 & + \frac13 g^3 M^{-6} f^{ABE}f^{BCF}f^{CAG} \mathscr{N}^E \mathscr{N}^F \mathscr{N}^G
\nonumber\\
 & - g^3 M^{-4}  f^{ABE}f^{BCF}f^{CAG} \mathscr{N}^E i\mathscr{\bar C}^F \mathscr{C}^G  + O(g^4) \Big]
  ,
  \label{J_2}
	\end{align}
where we have used
$
f^{AEC}f^{AEF}=N_c \delta^{CF}
$
Therefore, the correction from the measure to the Lagrangian density begins with the order $g^2$.
In other words, $J=1$ up to the order $g$.

This is also checked by using the relationship up to the order $g$ (\ref{A_mu-expan}):
\begin{equation}
\frac{\delta \mathscr{A}_{\mu}^{A}(x)}{\delta \mathscr{K}_{\nu}^B(y)} = \delta_\mu^\nu (\delta^{AB} + g M^{-2}f^{ABC}\mathscr{N}^C ) \delta^D(x-y) + O(g^2) .
\end{equation}
Thus, the Lagrangian density $\mathscr{L}_1=\mathscr{L}_1[\mathscr{K},\mathscr{C},\mathscr{\bar C},\mathscr{N}]$ up to the order $g$ does not change and remains in the form (\ref{L1}), and the calculation obtained using $\mathscr{L}_1$ (\ref{L1}) is not affected.

\section{Lagrangian and Feynman rules in the order $g^2$}

The relation $\mathscr{A}= \mathscr{\tilde A}[\mathscr{K}_\mu, \mathscr{C}, \bar{\mathscr{C}}, \mathscr{N}]$ up to the order $g^2$ is obtained as
\begin{align}
		\mathscr{A}_{\mu} =& \mathscr{K}_{\mu}
		+ \frac{1}{M^2} \partial_{\mu} \mathscr{N} - \frac{g}{M^2} i\bar{\mathscr{C}} \times \partial_{\mu}\mathscr{C}  \nonumber\\
		&
		+ \frac{g}{M^2} \mathscr{K}_{\mu} \times \mathscr{N}
		+ \frac{g}{M^4} \partial_{\mu} \mathscr{N} \times \mathscr{N} 
		\nonumber\\
		&- \frac{g^2}{M^4} (i\mathscr{\bar C} \times \partial_{\mu}\mathscr{C}) \times \mathscr{N} +   \frac{g^2}{M^4} (\mathscr{K}_\mu \times\mathscr{N}) \times \mathscr{N}  
\nonumber\\
		&+   \frac{g^2}{M^6} (\partial_\mu \mathscr{N} \times \mathscr{N}) \times \mathscr{N} - \frac{g^2}{M^2} i\mathscr{\bar C} \times (\mathscr{K}_\mu \times \mathscr{C}) 
\nonumber\\
		&-  \frac{g^2}{M^4} i\mathscr{\bar C} \times (\partial_\mu \mathscr{N} \times \mathscr{C}) 
+ O(g^3) .
		\label{A_mu-expan2}
\end{align}
The transformation (\ref{A_mu-expan2}) leads to 
\begin{align}
\frac{\delta \mathscr{A}_{\mu}^{A}(x)}{\delta \mathscr{K}_{\nu}^B(y)} 
= & \delta_\mu^\nu   (\delta^{AB} + g M^{-2}f^{ABC}\mathscr{N}^C  \nonumber\\
		&- g^2M^{-4}f^{ACE}f^{BFE} \mathscr{N}^C \mathscr{N}^F 
\nonumber\\&
- g^2M^{-2}f^{ACE}f^{BFE} i\bar{\mathscr{C}}^C \mathscr{C}^F) \delta^D(x-y)  \nonumber\\
		&+ O(g^3) .
\end{align}
Indeed, this gives the same Jacobian as (\ref{J_2}) up to the order $g^2$. 
In order to check the consistency of the Jacobian (\ref{J_2}), therefore, we must obtain the Lagrangian density $\mathscr{L}_2$ in the order $g^2$.

We must take into account the correction coming from the Jacobian $J$ in obtaining the Lagrangian density $\mathscr{L}_2=\mathscr{L}_2[\mathscr{K},\mathscr{C},\mathscr{\bar C},\mathscr{N}]$ in the order $g^2$.
Then the Lagrangian density $\mathscr{L}_2$ in the order $g^2$ is obtained as follows.
	\begin{align}
	\mathscr{L}_2 =& \sum_{*=a,b,c,d,e,f,g,h} \mathscr{L}_2^{(*)} - i \ln{J(g^2)} ,
	\nonumber\\
		\mathscr{L}_2^{(a)} =& - \frac{g^2}{4} (\mathscr{K}_{\mu} \times \mathscr{K}_{\nu})^2 ,
\nonumber
	\end{align}
	\begin{align}	
	\mathscr{L}_2^{(b)} =&+ \frac{g^2}{2M^4} (\partial_{\mu} \mathscr{K}_{\nu} \times \mathscr{N})^2 \nonumber\\
		&- \frac{g^2}{2M^4} (\partial_{\mu} \mathscr{K}_{\nu} \times \mathscr{N}) \cdot (\partial^{\nu} \mathscr{K}^{\mu} \times \mathscr{N}) \nonumber\\
		&+ \frac{g^2}{2M^4} (\mathscr{K}_{\mu} \times \mathscr{K}_{\nu}) \cdot (\partial^{\mu} \mathscr{N} \times \partial^{\nu} \mathscr{N}) \nonumber\\
		&- g^2 M^{-4} (\mathscr{K}_{\mu} \times \mathscr{N}) \cdot (\partial^{\mu} \mathscr{K}^{\nu} \times \partial_{\nu} \mathscr{N}) \nonumber\\
		&+ g^2 M^{-4} (\mathscr{K}_{\mu} \times \mathscr{N}) \cdot (\partial^{\nu} \mathscr{K}^{\mu} \times \partial_{\nu} \mathscr{N}) \nonumber\\
		&- \frac{g^2}{2M^2} (\mathscr{K}^{\mu} \times \mathscr{N})^2 ,
\nonumber
	\end{align}
	\begin{align}	
	\mathscr{L}_2^{(c)} =&+ g^2 M^{-6} (\partial^{\nu} \mathscr{K}^{\mu} \times \mathscr{N}) \cdot (\partial_{\mu} \mathscr{N} \times \partial_{\nu} \mathscr{N}) ,
\nonumber
	\end{align}
	\begin{align}	
\mathscr{L}_2^{(d)} =&- \frac{g^2}{4M^8} (\partial_{\mu} \mathscr{N} \times \partial_{\nu} \mathscr{N})^2 
		+ \frac{g^2}{2M^6} (\partial_{\mu} \mathscr{N} \times \mathscr{N})^2 ,
\nonumber
	\end{align}
	\begin{align}
		\mathscr{L}_2^{(e)} =&- g^2 M^{-2} (\partial^{\mu} \partial_{\mu} \mathscr{K}^{\nu} \times i \bar{\mathscr{C}}) \cdot (\mathscr{K}_{\nu} \times \mathscr{C}) \nonumber\\
		&+ g^2 M^{-2} (\partial^{\nu} \partial_{\mu} \mathscr{K}^{\mu} \times i \bar{\mathscr{C}}) \cdot (\mathscr{K}_{\nu} \times \mathscr{C}) \nonumber\\
		&+ g^2 M^{-2} (\partial^{\mu} \mathscr{K}^{\nu} \times \mathscr{K}_{\mu}) \cdot (i \bar{\mathscr{C}} \times \partial_{\nu} \mathscr{C}) \nonumber\\
		&- g^2 M^{-2} (\partial^{\nu} \mathscr{K}^{\mu} \times \mathscr{K}_{\mu}) \cdot (i \bar{\mathscr{C}} \times \partial_{\nu} \mathscr{C}) \nonumber\\
		&+ g^2 M^{-2} (\mathscr{K}^{\mu} \times \mathscr{K}^{\nu}) \cdot (i \partial_{\mu} \bar{\mathscr{C}} \times \partial_{\nu} \mathscr{C}) \nonumber\\		
		&+ g^2 (\mathscr{K}_{\mu} \times \mathscr{C}) \cdot ( \mathscr{K}^{\mu} \times i \bar{\mathscr{C}}) ,
\nonumber
\\
   \mathscr{L}_2^{(f)} =&- g^2 M^{-4} (\partial_{\mu} \partial^{\mu} \mathscr{K}^{\nu} \times i \bar{\mathscr{C}}) \cdot (\partial_{\nu} \mathscr{N} \times \mathscr{C}) \nonumber\\
		&+ g^2 M^{-4} (\partial_{\mu} \partial^{\nu} \mathscr{K}^{\mu} \times i \bar{\mathscr{C}}) \cdot (\partial_{\nu} \mathscr{N} \times \mathscr{C}) \nonumber\\
		&+ g^2 M^{-2} (\partial_{\mu} \mathscr{N} \times \mathscr{C}) \cdot (\mathscr{K}^{\mu} \times i \bar{\mathscr{C}}) \nonumber\\
		&+ g^2 M^{-2} (\mathscr{K}^{\mu} \times \mathscr{N}) \cdot (i \partial_{\mu} \bar{\mathscr{C}} \times \mathscr{C}) ,
\nonumber
\\
		\mathscr{L}_2^{(g)} =&- \frac{g^2}{2M^6} (i \partial^{\mu} \bar{\mathscr{C}} \times \partial^{\nu} \mathscr{C}) \cdot (\partial_{\mu} \mathscr{N} \times \partial_{\nu} \mathscr{N}) \nonumber\\
		&+ g^2 M^{-4} (\partial_{\mu} \mathscr{N} \times \mathscr{N}) \cdot (i \partial^{\mu} \bar{\mathscr{C}} \times \mathscr{C}) \nonumber\\
		&- g^2 M^{-4} (\partial_{\mu} \mathscr{N} \times \mathscr{N}) \cdot (i \bar{\mathscr{C}} \times \partial^{\mu} \mathscr{C}) ,
\nonumber
	\end{align}
	\begin{align}	
     \mathscr{L}_2^{(h)} =&- \frac{g^2}{2M^4} (i \partial^{\mu} \bar{\mathscr{C}} \times \partial^{\nu} \mathscr{C})^2
		+ \frac{g^2}{2M^2} (i \bar{\mathscr{C}} \times \partial_{\mu} \mathscr{C})^2 \nonumber\\
		&+ \frac{g^2}{2M^4} (i \partial^{\mu} \bar{\mathscr{C}} \times \partial^{\nu} \mathscr{C}) \cdot (i \partial_{\nu} \bar{\mathscr{C}} \times \partial_{\mu} \mathscr{C}) \nonumber\\
		&- \frac{g^2}{M^2} (i \bar{\mathscr{C}} \times \partial_{\mu} \mathscr{C}) \cdot (i \partial^{\mu} \bar{\mathscr{C}} \times \mathscr{C}) \nonumber\\
		&+ \frac{\beta}{4} (i \bar{\mathscr{C}} \times \mathscr{C})^2 . 
 %\nonumber\\
%		&- i \ln{J(g^2)} .
		\label{L2}
	\end{align}

The following are the Feynman rules for the vertex functions of the order $g^2$ obtained from (\ref{L2}).

\renewcommand{\theenumi}{\alph{enumi}}
\renewcommand{\labelenumi}{(\theenumi)}
\begin{enumerate}
\item $\mathscr{K} \mathscr{K} \mathscr{K} \mathscr{K}$ vertex:
	\begin{align}
		&\langle \mathscr{K}^{A}_{\mu}(k_1) \mathscr{K}^{B}_{\nu}(k_2) \mathscr{K}^{C}_{\rho}(k_3) \mathscr{K}^{D}_{\sigma}(k_4) \rangle \nonumber\\
		=& - g^2 W^{ABCD}_{\mu \nu \rho \sigma} ,
	\end{align}
with
	\begin{align}
		W^{ABCD}_{\mu \nu \rho \sigma} =& (f^{AC,BD}-f^{AD,CB})g_{\mu\nu}g_{\rho\sigma} \nonumber\\
		&+ (f^{AB,CD}-f^{AD,BC})g_{\mu\rho}g_{\nu\sigma} \nonumber\\
		&+ (f^{AC,DB}-f^{AB,CD})g_{\mu\sigma}g_{\nu\rho}, \nonumber\\
		f^{AB,CD} =& f^{ABE}f^{CDE} .
	\end{align}

\item $\mathscr{K} \mathscr{K} \mathscr{N} \mathscr{N}$ vertex:
	\begin{align}
		&\langle \mathscr{K}^{A}_{\mu}(k_1) \mathscr{K}^{B}_{\nu}(k_2) \mathscr{N}^{C}(k_3) \mathscr{N}^{D}(k_4) \rangle \nonumber\\
		=& - \frac{g^2}{M^2} g_{\mu\nu} (f^{AC,BD} + f^{AD,BC}) \nonumber\\
		& - \frac{g^2}{M^4} g_{\mu\nu} \big[ (k_1 \cdot k_3 + k_2 \cdot k_4 + k_1 \cdot k_2) f^{AC,BD} \nonumber\\
		& \hspace{50pt} + (k_1 \cdot k_4 + k_2 \cdot k_3 + k_1 \cdot k_2) f^{AD,BC} \big] \nonumber\\
		& + \frac{g^2}{M^4} \big[ (k_{1\nu} k_{2\mu} + k_{1\nu} k_{3\mu} + k_{4\nu} k_{2\mu}) f^{AC,BD} \nonumber\\
		& \hspace{30pt} + (k_{1\nu} k_{2\mu} + k_{1\nu} k_{4\mu} + k_{3\nu} k_{2\mu}) f^{AD,BC} \big] \nonumber\\
		& + \frac{g^2}{M^4} f^{AB,CD} (k_{3\nu} k_{4\mu} - k_{3\mu} k_{4\nu}) .
	\end{align}

\item $\mathscr{K} \mathscr{N} \mathscr{N} \mathscr{N}$ vertex:
	\begin{align}
		&\langle \mathscr{K}^{A}_{\mu}(k_1) \mathscr{N}^{B}(k_2) \mathscr{N}^{C}(k_3) \mathscr{N}^{D}(k_4) \rangle  \nonumber\\
		&= \frac{ig^2}{M^6} \Big[ (k_1 \cdot k_2) (f^{AD,BC} k_{3\mu} + f^{AC,BD} k_{4\mu}) \nonumber\\
		& \hspace{25pt} + (k_1 \cdot k_3) (f^{AD,CB} k_{2\mu} + f^{AB,CD} k_{4\mu}) \nonumber\\
		& \hspace{25pt} + (k_1 \cdot k_4) (f^{AC,DB} k_{2\mu} + f^{AB,DC} k_{3\mu}) \Big] .
	\end{align}
	
\item $\mathscr{N} \mathscr{N} \mathscr{N} \mathscr{N}$ vertex:
	\begin{align}
		&\langle \mathscr{N}^{A}(k_1) \mathscr{N}^{B}(k_2) \mathscr{N}^{C}(k_3) \mathscr{N}^{D}(k_4) \rangle \nonumber\\
		=& - \frac{g^2}{M^8} k_1^{\mu} k_2^{\nu} k_3^{\rho} k_4^{\sigma} W^{ABCD}_{\mu\nu\rho\sigma} \nonumber\\
		& - \frac{g^2}{M^6} \Big\{ (k_1 \cdot k_2 + k_3 \cdot k_4) (f^{AC,BD} + f^{AD,BC}) \nonumber\\
		& \qquad + (k_1 \cdot k_3 + k_2 \cdot k_4) (f^{AB,CD} + f^{AD,CB}) \nonumber\\
		& \qquad + (k_1 \cdot k_4 + k_2 \cdot k_3) (f^{AB,DC} + f^{AC,DB}) \Big\} .
	\end{align}

\item $\mathscr{K} \mathscr{K} \mathscr{C} \bar{\mathscr{C}}$ vertex:
	\begin{align}
		&\langle \mathscr{K}^{A}_{\mu}(k_1) \mathscr{K}^{B}_{\nu}(k_2) \bar{\mathscr{C}}^{C}(k_3) \mathscr{C}^{D}(k_4) \rangle \nonumber\\
		=& - M^2 \langle \mathscr{K}^{A}_{\mu}(k_1) \mathscr{K}^{B}_{\nu}(k_2) \mathscr{N}^{C}(k_3) \mathscr{N}^{D}(k_4) \rangle . 
	\end{align}

\item $\mathscr{K} \mathscr{N} \mathscr{C} \bar{\mathscr{C}}$ vertex:
	\begin{align}
		&\langle \mathscr{K}^{A}_{\mu}(k_1) \mathscr{N}^{B}(k_2) \bar{\mathscr{C}}^{C}(k_3) \mathscr{C}^{D}(k_4) \rangle \nonumber\\
		=& \frac{ig^2}{M^2} (f^{AB,CD} k_{3\mu} - f^{AC,BD} k_{2\mu}) \nonumber\\
		&+ \frac{ig^2}{M^4} [k_1^2 k_{2\mu} - (k_1 \cdot k_2) k_{1\mu}] f^{AC,BD} .
	\end{align}
	
\item $\mathscr{N} \mathscr{N} \mathscr{C} \bar{\mathscr{C}}$ vertex:
	\begin{align}
		&\langle \mathscr{N}^{A}(k_1) \mathscr{N}^{B}(k_2) \bar{\mathscr{C}}^{C}(k_3) \mathscr{C}^{D}(k_4) \rangle \nonumber\\
		=& \frac{g^2}{M^4} [(k_1 \cdot k_4) + (k_2 \cdot k_3) - (k_1 \cdot k_3) - (k_2 \cdot k_4)] f^{AB,CD} \nonumber\\
		&+ \frac{g^2}{2M^6} [(k_1 \cdot k_4) (k_2 \cdot k_3) - (k_1 \cdot k_3) (k_2 \cdot k_4)] f^{AB,CD} .
	\end{align}
	
\item $\mathscr{C} \bar{\mathscr{C}} \mathscr{C} \bar{\mathscr{C}}$ vertex:
	\begin{align}
		&\langle \bar{\mathscr{C}}^{A}(k_1) \mathscr{C}^{B}(k_2) \bar{\mathscr{C}}^{C}(k_3) \mathscr{C}^{D}(k_4) \rangle \nonumber\\
		=& (f^{AB,CD} + f^{AD,BC}) \nonumber\\
		& \times \left[ \frac{\beta}{2} g^2 - \frac{g^2}{M^2} (k_2 \cdot k_4) - \frac{g^2}{M^4} (k_1 \cdot k_3) (k_2 \cdot k_4) \right] \nonumber\\
		& + \frac{g^2}{M^2} \big[ (k_1 \cdot k_4 + k_2 \cdot k_3) f^{AB,CD} \nonumber\\
		& \hspace{30pt} + (k_1 \cdot k_2 + k_3 \cdot k_4) f^{AD,BC} \big] \nonumber\\
		& + \frac{g^2}{M^4} \big[ (k_1 \cdot k_4) (k_2 \cdot k_3) f^{AB,CD} \nonumber\\
		& \hspace{30pt} + (k_1 \cdot k_2) (k_3 \cdot k_4) f^{AD,BC} \big] .
	\end{align}
	
\end{enumerate}

%%%%%%%%%%   REFERENCES   %%%%%%%%%%


\begin{thebibliography}{99}
\bibitem{YM54}
C.N. Yang and R.L. Mills,
%Conservation of isotopic spin and isotopic gauge invariance,
Phys. Rev. {\bf 96}, 191%--195
 (1954).


\bibitem{tHooft71}
G. 't Hooft,
%Renormalization of Massless Yang-Mills Fields, 
Nucl. Phys. B{\bf 33}, 173%--199
 (1971).


\bibitem{tHooft71b}
G. 't Hooft,
%Renormalizable Lagrangians for massive Yang-Mills fields,
Nucl. Phys. B{\bf 35}, 167%--188
 (1971).


\bibitem{Higgs66}
P.W. Higgs,
%Spontaneous symmetry breakdown without massless bosons,
Phys. Rev. {\bf 145}, 1156%--1163
 (1966).


%\bibitem{noHiggsmodel}
\bibitem{DV70}
H. van Dam and M.J.G. Veltman,
%Massive and massless Yang-Mills and gravitational fields,
Nucl. Phys. B{\bf 22}, 397%--411
 (1970) 


\bibitem{SF70}
A.A. Slavnov and L.D. Faddeev, 
%Massless and massive yang-mills field,
Theor. Math. Phys. {\bf 3},  312%--316
 (1971) [Teor. Mat. Fiz. {\bf 3}, 18%--23
 (1970)]. 
\\
A.A. Slavnov,
%Massive gauge fields,
Theor. Math. Phys. {\bf 10},  201%--217
 (1972) [Teor. Mat. Fiz. {\bf 10}, 305%--328
 (1972)]. 


\bibitem{Boulware70}
D.G. Boulware, 
%Renormalizeability of massive non-abelian gauge fields - a functional integral approach,
Annals Phys. {\bf 56}, 140%--171
 (1970). 


\bibitem{CF76}
G. Curci and R. Ferrari,
%On a Class of Lagrangian Models for Massive and Massless Yang-Mills Fields,
Nuovo Cim. A{\bf 32}, 151%--168
 (1976).


\bibitem{CF76b}
G. Curci and R. Ferrari, 
%The Unitarity Problem and the Zero-Mass Limit for a Model of Massive Yang-Mills Theory,
Nuovo Cim. A{\bf 35}, 1%--14
 (1976), Erratum-ibid. A{\bf 47}, 555 (1978). 


\bibitem{Ojima82}
I. Ojima, 
%Comments On Massive And Massless Yang-mills Lagrangians With A Quartic Coupling Of Faddeev-popov Ghosts,
Z. Phys. C{\bf 13}, 173%--177
 (1982).


\bibitem{BSNW96}
J. de Boer, K. Skenderis, P. van Nieuwenhuizen and A. Waldron,
%On the renormalizability and unitarity of the Curci-Ferrari model for massive vector bosons,
%e-Print: hep-th/9510167, 
Phys. Lett. B{\bf 367},  175%--182
 (1996).
 

\bibitem{FMTY81}
T. Kunimasa and T. Goto,  
%Generalization of the Stueckelberg Formalism to the Massive Yang-Mills Field,
Prog. Theor. Phys. {\bf 37}, 452%--464
 (1967). 
\\
T. Fukuda, M. Monda, M. Takeda and Kan-ichi Yokoyama,
%Quantum Theory Of Massive Yang-mills Fields. 1. Basis Of Formulation With Indefinite Metric,
Prog. Theor. Phys. {\bf 66}, 1827%--1842
 (1981).
\\
J.M. Cornwall, 
%Dynamical Mass Generation in Continuum QCD,
Phys. Rev. D{\bf 26}, 1453%--1478
 (1982). 


\bibitem{DTT88}
R. Delbourgo, S. Twisk and G. Thompson,
%Massive Yang-Mills theory: renormalizability and unitarity,
Int. J. Mod. Phys. A{\bf 3}, 435%--449
 (1988).


\bibitem{RRA04}
H. Ruegg and M. Ruiz-Altaba,
%The Stuckelberg field.
%e-Print: hep-th/0304245,
Int. J. Mod. Phys. A{\bf 19},  3265%--3348
 (2004).


\bibitem{BFQ}
R. Ferrari,   
%Metamorphosis versus Decoupling in Nonabelian Gauge Theories at Very High Energies,
arXiv:1106.5537 [hep-ph],
Acta Phys.Polon. B\textbf{43}, 1735-1767  (2012).
\\
D. Bettinelli, R. Ferrari and A. Quadri, 
%One-loop Self-energies in the Electroweak Model with Nonlinearly Realized Gauge Group,
arXiv:0903.0281 [hep-th], 
Phys. Rev. D79, 125028 (2009), 
Erratum-ibid. D85, 049903 (2012).
\\
D. Bettinelli, R. Ferrari and A. Quadri, 
%The SU(2) x U(1) Electroweak Model based on the Nonlinearly Realized Gauge Group. II. Functional Equations and the Weak Power-Counting,
arXiv:0809.1994 [hep-th], 
Acta Phys. Polon B41, 597--628 (2010).
\\
D. Bettinelli, R. Ferrari and A. Quadri, 
%One-loop self-energy and counterterms in a massive Yang-Mills theory based on the nonlinearly realized gauge group,
arXiv:0709.0644 [hep-th],  
Phys. Rev. D77, 105012 (2008).
\\
D. Bettinelli, R. Ferrari and A. Quadri, 
%A Massive Yang-Mills Theory based on the Nonlinearly Realized Gauge Group,
arXiv:0705.2339 [hep-th], 
Phys. Rev. D77, 045021 (2008).
\\
R. Ferrari and A. Quadri, 
%Physical unitarity for massive non-Abelian gauge theories in the Landau gauge: Stueckelberg and Higgs,
hep-th/0408168, 
JHEP 0411, 019 (2004).


\bibitem{Cornwall82}
J.M. Cornwall and A. Soni,  
%Glueballs as Bound States of Massive Gluons,
Phys. Lett. B{\bf 120}, 431%--435
 (1983). 


\bibitem{decoupling}
 Ph. Boucaud, J.P. Leroy, A. Le Yaouanc, J. Micheli, O. Pene and J. Rodriguez-Quintero, 
%On the IR behavier of the Landau-gauge ghost propagator,
[hep-ph/0803.2161],
JHEP 0806, 099 (2008).
%\\
% Ph. Boucaud, J.P. Leroy, A. Le Yaouanc, J. Micheli, O. Pene and J. Rodriguez-Quintero, 
%IR finiteness of the ghost dressing function from numerical resolution of the ghost SD equation,
%arXiv:0801.2721[hep-ph],
%JHEP 0806, 012 (2008).
\\
%\bibitem{ABP08}
 A.C. Aguilar, D. Binosi and J. Papavassiliou,
%Gluon and ghost propagators in the Landau gauge: Deriving lattice results from Schwinger-Dyson equations,
arXiv:0802.1870 [hep-ph],
Phys. Rev. D{\bf 78}, 025010 (2008).
\\ 
%\bibitem{FMP09}
C.S.~Fischer, A.~Maas and J.M.~Pawlowski,
%On the infrared behavior of Landau gauge Yang-Mills theory,
arXiv:0810.1987 [hep-ph],
Annals Phys.\textbf{324}, 2408%--2437
 (2009).
\\
%\bibitem{BGP10}
J. Braun, H. Gies and J.M. Pawlowski, 
%Quark Confinement from Color Confinement,
arXiv:0708.2413 [hep-th], 
Phys. Lett. B{\bf 684}, 262%--267
 (2010). 


\bibitem{scaling}
%\bibitem{AS01}
R. Alkofer and L. von Smekal, 
%The Infrared behavior of QCD Green's functions: Confinement dynamical symmetry breaking, and hadrons as relativistic bound states,
[hep-ph/0007355], 
Phys. Rep. \textbf{353}, 281%--465
 (2001).


\bibitem{decoupling-lattice}
%\bibitem{BIMPS07}
I.L. Bogolubsky, E.M. Ilgenfritz, M. Muller-Preussker and A. Sternbeck, 
%The Landau gauge gluon and ghost propagators in 4D SU(3) gluodynamics in large lattice volumes,
arXiv:0710.1968 [hep-lat], 
PoS LAT2007, 290 (2007). 
\\
A. Cucchieri and T. Mendes,
%What's up with IR gluon and ghost propagators in Landau gauge? A puzzling answer from huge lattices, 
arXiv:0710.0412 [hep-lat], 
PoS LAT2007, 297 (2007). 
\\
A. Sternbeck, L. von Smekal, D.B. Leinweber  and A.G. Williams, 
%Comparing SU(2) to SU(3) gluodynamics on large lattices,
arXiv:0710.1982 [hep-lat] 
PoS LAT2007, 340 (2007). 



\bibitem{TW11}
M. Tissier and N. Wschebor,  
%An Infrared Safe perturbative approach to Yang-Mills correlators,
arXiv:1105.2475 [hep-th], 
Phys. Rev. D\textbf{84}, 045018 (2011). 
\\
M. Tissier and N. Wschebor,  
%Infrared propagators of Yang-Mills theory from perturbation theory, 
arXiv:1004.1607 [hep-ph], 
Phys. Rev. D\textbf{82}, 101701 (2010). 
\\
J. Serreau and M. Tissier,
%Lifting the Gribov ambiguity in Yang-Mills theories,
arXiv:1202.3432 [hep-th], 
Phys.Lett. B\textbf{712},  97--103 (2012).


\bibitem{Baulieu85}
L. Baulieu,  
%Perturbative Gauge Theories,
Phys. Rept. {\bf 129}, 1%--74
 (1985).


\bibitem{Kondo12}
K.-I. Kondo,
%A nonperturbative construction of massive Yang-Mills fields without Higgs fields,
arXiv:1208.3521[hep-th],
Phys. Rev. D\textbf{87}, 025008 (2013).
  

\bibitem{KO79}
 T. Kugo and I. Ojima,
%Local covariant operator formalism of non-Abelian gauge theories and quark confinement problem,
Suppl. Prog. Theor. Phys. \textbf{66}, 1%--130 
(1979).
%\\
%T. Kugo,
%The universal renormalization factors $Z_1/Z_3$ and color confinement condition in non-Abelian gauge theory, 
%hep-th/9511033.


\bibitem{KO78}
T. Kugo and I. Ojima,
%Manifestly Covariant Canonical Formulation of Yang-Mills Field Theories: Physical State Subsidiary Conditions and Physical S Matrix Unitarity,
Phys. Lett. B{\bf 73}, 459%--462
 (1978).


\bibitem{Gribov78}
V.N. Gribov. 
%Quantization of Nonabelian Gauge Theories,
Nucl Phys. B\textbf{139}, 1%--19
 (1978).


\bibitem{Kondo12a}
%K.-I. Kondo,
%A unitary and renormalizable model for massive Yang-Mills fields without Higgs fields.
%arXiv:1202.4162 [hep-th], 
%to be revised.
K.-I. Kondo et al.,
 papers in preparation.


\bibitem{PS95}
 M. E. Peskin and D.V. Schroeder,
An Introduction to quantum field theory 
(Addison-Wesley, Reading, USA, 1995).

 
\bibitem{LQT77}
B.W. Lee, C. Quigg, H.B. Thacker  
%Weak Interactions at Very High-Energies: The Role of the Higgs Boson Mass,
Phys. Rev. D\textbf{16},  1519%--1531
 (1977). 


\bibitem{CLT73}
J.M. Cornwall, D.N. Levin and G. Tiktopoulos,
%Uniqueness of spontaneously broken gauge theories,
Phys. Rev. Lett. \textbf{30}, 1268%--1270
  (1973), 
Erratum-ibid. \textbf{31}, 572 (1973). 
\\
%Derivation of Gauge Invariance from High-Energy Unitarity Bounds on the s Matrix.
Phys. Rev. D\textbf{10}, 1145%--1167
 (1974),
Erratum-ibid. D\textbf{11}, 972 (1975). 


\bibitem{Nakanishi72}
N. Nakanishi,
%Massive vector field and electromagnetic field in the landau gauge,
Phys. Rev. D{\bf 5},  1324%--1330
 (1972).

 
\bibitem{Lavrov12}
P.M. Lavrov,
%Remarks on the Curci-Ferrari model,
arXiv:1205.0620 [hep-th], 
Mod.Phys.Lett. A\textbf{27}, 1250132 (2012). 


\bibitem{Kondo01}
K.-I. Kondo,
%Vacuum condensate of mass dimension 2 as the origin of mass gap and quark confinement,
[hep-th/0105299], 
Phys. Lett. B\textbf{514}, 335%--345
 (2001). 
\\
K.-I. Kondo,
%A Physical meaning of mixed gluon ghost condensate of mass dimension two.
[hep-th/0306195], 
Phys. Lett. B\textbf{572}, 210%--215
 (2003). 


\bibitem{KMSI02}
K.-I. Kondo, T. Murakami, T. Shinohara and T. Imai, 
%Renormalizing a BRST invariant composite operator of mass dimension 2 in Yang-Mills theory,
[hep-th/0111256], 
Phys. Rev. D\textbf{65}, 085034 (2002). 


\bibitem{Cutkosky60}
R.E. Cutkosky,
%Singularities and discontinuities of Feynman amplitudes,
J. Math. Phys. {\bf 1}, 429%--433
 (1960).




\end{thebibliography}
\end{document}